\documentclass[a4paper,10pt,notitlepage]{revtex4}
\usepackage{normal}
\usepackage{datetime}
\usepackage{graphicx}
\RequirePackage{amsmath}
\RequirePackage{mathtools} 
\usepackage{amsfonts}
\usepackage{amssymb}
\usepackage{dsfont}
\usepackage{latexsym}
\usepackage{url}
\usepackage{subfigure}
\usepackage{wrapfig}
\usepackage{tikz}
\usetikzlibrary{shapes,arrows}
\tikzstyle{every loop}=[]
\newcommand{\etal}{et.~al.~}
\newcommand{\ie}{i.e.~}

\def\C{{\cal C}}
\newcommand{\ket}[1]{\ensuremath{\left|#1\right\rangle}}

\newcommand{\p}{\mathbf{p}}
\newcommand{\RT}{\mathcal{R_T}}
\newcommand{\IRT}{\mathcal{\widetilde{R}_T}}
\newcommand{\theory}{\mathcal{T}}
\newcommand{\see}{C}

\begin{document}
\title
{When does a physical system compute?}
\author
{%
Dominic Horsman$^{1}$,
Susan Stepney$^{2}$,
Rob C.~Wagner$^{3}$
and Viv Kendon$^{3}$%
}

\address{%
$^{1}$Department of Computer Science, University of Oxford, Oxford, OX1 3QD, UK\\
$^{2}$Department of Computer Science, and York Centre for Complex Systems Analysis, University of York, York, YO10 5GH, UK\\
$^{3}$School of Physics and Astronomy, University of Leeds, Leeds, LS2 9JT, UK
}

%
%

\begin{abstract}

Computing is a high-level process of a physical system. Recent interest in non-standard computing systems, including quantum and biological computers, has brought this physical basis of computing to the forefront. There has been, however, no consensus on how to tell if a given physical system is acting as a computer or not; leading to confusion over novel computational devices, and even claims that every physical event is a computation. In this paper we introduce a formal framework that can be used to determine whether a physical system is performing a computation. We demonstrate how the abstract computational level interacts with the physical device level, in comparison to the use of mathematical models in experimental science. This powerful formulation allows a precise description of experiments, technology, computation, and simulation, giving our central conclusion: \emph{physical computing is the use of a physical system to predict the outcome of an abstract evolution}. We give conditions for computing, illustrated using a range of non-standard computing scenarios. The framework also covers broader computing contexts, where there is no obvious human computer user. We introduce the notion of a `computational entity', and its critical role in defining when computing is taking place in physical systems.  \\


\indent \emph{Keywords:} computation, physical computation, computer. 

\end{abstract}

\date{\today}
\label{firstpage}
\maketitle
%
\section{Introduction}\label{intro} 


Information science is one of the great advances of the last century. The technology that developed from it is now integral to almost all aspects of day-to-day life in the developed world, and advances in mobile telephone hardware have put a computer in (almost) every pocket. In addition to the proliferation of semiconductor-based computers, non-standard (also known as unconventional) computational systems continue to be proposed and used -- from the differential analysers of the early 20th century \cite{Shannon1941}, through to the recent explosion of interest in quantum computing \cite{Ladd2010,nandc}, and other proposals such as quantum annealing \cite{Johnson2011}, DNA \cite{dna1,dna2}, or chemical \cite{chem1,chem2} computational devices. The notion of computation, and its related system property, information, has been imported into other fields in an attempt to describe and explain such diverse processes as photosynthesis \cite{ball2011} and the conscious mind \cite{SOTM}, and a strand of modern cross-discipline thought has given us the claims that ``everything is information" \cite{vlatkobook} or ``the universe is a [quantum] computer" \cite{lloydqc}.

In parallel with the technological and conceptual development of information science, its foundations continue to be addressed. The definition of which mathematical, logical, and algorithmic structures constitute `a computation' is a topic of ongoing research (see for example \cite{comp,flc}). The question of how to define information, both as a concept and a physical quantity, is being investigated by philosophers, physicists, and informatics researchers (see for example \cite{sepinformation}). In this paper we address a third, equally important, and specifically physical, question: what is a computer? Given some notion of a mathematical computation, what does it mean to say that some physical system is `running' a computation? 
If we want to use computational notions in physics, what are the necessary and sufficient conditions under which we can say that a particular physical system is carrying out a computation? In short -- \emph{when does a physical system compute}?

There is currently no accepted answer to this question, and an absence of a worked out formalism within which to determine if a computation is happening physically gives rise to a great deal of confusion when discussing non-standard forms of computation. We can all agree that a laptop running a Matlab calculation and a server processing search engine queries are physical systems performing computation. However, when we move beyond standard and mass-produced technology, the question becomes more difficult to answer. Is a protein performing a compaction computation as it folds \cite{annurev}? Does a photon (quantum) compute the shortest path through a leaf in photosynthesis \cite{mohseni2008environment}? Is the human mind a computer \cite{ENM}? A dog catching a stick \cite{dogstick}? A stone sitting on the floor \cite{stone}? One answer is that they all are -- that everything that physically exists is performing computation by virtue of its existence. Unfortunately, by thus defining the universe and everything in it as a computer, the notion of physical computation becomes empty. To state that \emph{every} physical process is a computation is simply to re-define what is meant by a `physical process' -- there is then no nontrivial content to the assertion. A statement such as ``everything is computation" is either false, or it is trivial; either way, it is not useful in determining properties of physical systems in practice

In this paper we give a framework that can be used to determine whether a physical system is computing or not. We define what it means for a physical system to compute a mathematically-defined computation, how the physical and mathematical levels in computing interact, and give necessary conditions for a physical system to be computing. Key to defining our framework is the \emph{representation relation} that is fundamental in the physical sciences, where physical systems are represented by mathematical objects. We show how such a representation allows comparisons to be made between physical processes and mathematically-described computations, and how this can then be used to define when a physical process is being used in such a relation. This requires, explicitly, the notion of a `computational entity' to be necessary for a computation to proceed: we define such entities, show how such a definition does not require either intention or a conscious (or human) user, and argue that such a contextual notion of computation is not problematic for an account of a physically real process. In our framework, computation shares formal and structural similarities with scientific experiments and engineering technology: we are able to show precisely how they are related, and give exact definitions for each in terms of a single, underlying structure. In all cases, we are dealing with questions of representation: how is a physical system represented mathematically, how do we test that representation, and how can the representation be `reversed' so that a physical system can instantiate a mathematical description. As well as computation, these are key issues in how we determine between scientific theories by argument and experiment; and, in turn, fit into broader questions of representation that are fundamental to a number of different fields \cite{vanF}.

\section{Physical computation}



The question of when a physical system is computing is fundamentally a question about the relationship of abstract mathematical/logical entities to physical ones \cite{copeland}. A `computation' is a mathematical abstraction described in one of the logical formalisms developed by theoretical computer scientists.
A `computer' is a physical system with actual constituent parts and its own internal interactions that take it from one physical state to another. The computer is taken to stand in a certain relation to the computation -- if we can formulate this relation, then we can answer our question of when a physical system is performing computation. To act as a computer is always to be performing a specific computation; we therefore need to ask: when is \emph{this} physical system performing \emph{that} (not always known) computation, and what is the relation required between the physical system and the abstract computation that this can be determined?

The above gives us a view as in figure \ref{RR}(a): there is a space of abstract mathematical/logical entities and a space of physical entities. A computation is an entity in the first, and a putative computer in the second. So what is it that allows us to go between the two spaces? There is no possible notion of causation between them (this is simply a category error); so how does the abstract interact with the physical at all?

To answer this, we turn to the area where the question of the relation between abstract and physical has most commonly been posed: physics. Physics operates by representing physical systems abstractly, using abstract theory to predict the outcome of physical evolution, and formulating physical experiments to test the outcome of theoretical predictions. Physics works by constant and two-way interaction between abstract and physical. Exactly how it does this has been the subject of philosophical investigation for centuries, and while progress has been made, there is no clear and definitive description of the scientific process (see for example \cite{popper,kuhn,sciimage,french}). However, there are certain things that we can and cannot say about the specific question of the relationship between abstract description and physical entity. We use these to build a framework in which outstanding questions can be located, and which enables us to use what is known about the process of physics to show the relationship between physics and computing; and thereby to describe physical computation. It is important to note here that we are \emph{not} claiming to solve the problems of the philosophy of science. The framework we propose will hopefully be of interest to people in this field, but it has been constructed with the aim not of solving current issues but rather re-describing them.

\section{Physics and the representation relation}\label{physandrr}

The key to the interaction between abstract and physical entities in physics is via the \emph{representation relation} (see for example \cite{vanF,frigg}). This is the method by which physical systems are given abstract descriptions: an atom is represented as a wavefunction, a billiard ball as a point in phase space, a black hole as a metric tensor, and so on. That this relation is possible is a pre-requisite for physics: without a way of describing objects abstractly, we cannot do science. We have given examples of mathematical representation, but this is not necessary: it can be any abstract description of an object, logical, mathematical, or linguistic. Which type of representation has an impact on what sort of physics is possible: if we have a linguistic representation of object weight that is simply ``heavy'' or ``light'', then we are able to do much less precise physics than if we use a numerical amount of newtons. 

\begin{figure}[t]
    \scalebox{0.75}{\hspace{-2cm}

 \begin{minipage}[c]{1.0\linewidth}
  \[
   \begin{array}{ccc}
   \begin{tikzpicture}[font=\large]

\draw[style=dashed] (0,0) -- (6,0);
\draw (0,0.5) node {$Abstract$};
\draw (0,-0.5) node {$Physical$};

\node[circle,draw] at (3,-1.5) {$e^-$};
\draw (4,1.25) node {$\psi: \ i \hbar \frac{\partial \psi}{\partial t} = H \psi$};

\end{tikzpicture}
& \qquad \qquad \qquad
\quad &
\begin{tikzpicture}[font=\large]

\draw[style=dashed] (0,0) -- (6,0);
\draw (0,0.5) node {$Abstract$};
\draw (0,-0.5) node {$Physical$};

\node[circle,draw] at (3,-1.5) {$e^-$};
\draw (4,1.25) node {$\psi: \ i \hbar \frac{\partial \psi}{\partial t} = H \psi$};

\draw[->,double] (3,-1) -- (3,1);
\draw (3.5,0.25) node {$\mathcal{R}$};

\end{tikzpicture}
\\{}\\
 \mathrm{(a)} & \qquad \quad  &  \mathrm{(b)}  \end{array}
\]
\end{minipage}}
\caption{Representation in physics. (a) Spaces of abstract and physical objects (here, an electron and a wavefunction). (b) The representation relation used as the modelling relation $\mathcal{R}$ mediating between the spaces.} \setcounter{figure}{1}\label{RR}
\end{figure}
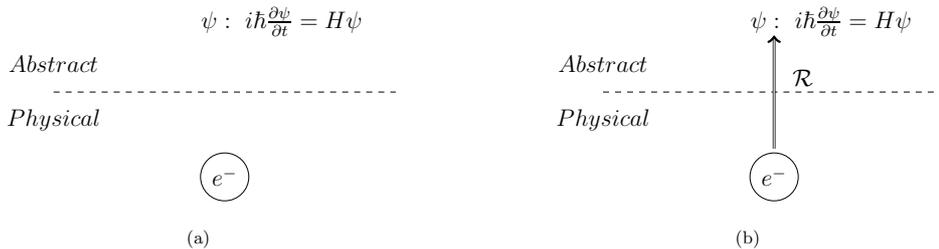

The most important property of the representation relation is that it is the relation that takes us across the divide between abstract and physical. The representation relation is unique in this respect, allowing a map between physical and abstract spaces: when we represent the physical and abstract as in figure \ref{RR} (and subsequent figures), we are referring to the spaces themselves, not mathematical descriptions of them, and the representation relation is not a mathematical relation. Precisely what it is, how it exists (and indeed can possibly exist) is a matter of ongoing research for philosophers of science; we know, nevertheless, that such a thing does exist. The representation relation is the relation that allows us to deal with the physical world at an abstract level; without it, any abstract reasoning about the physical world is not possible.

For a physicist, there is very little mystery in the representation relation: it is how physics works. This relation is how we can write down $\ket{\psi}$ and think that we are talking about an electron or a hydrogen atom or a Bose-Einstein condensate. Every time we use something abstract to represent something physical, we use a representation relation. It is important to notice that the representation of any given system is not unique: for example, a rubidium atom can be represented as a quantum bit (qubit), or as the solution to a master equation, or as a multi-level system with many orbitals. 

This initial use of the representation relation in physics is fundamentally the process of \emph{modelling}: an electron is modelled as a wavefunction, an aeroplane as a vector, and so on \cite[part1]{vanF}. The modelling relation $\mathcal{R}$ takes an individual physical entity $\p$ to its abstract model $m_\p$. We use lower case for individual entities and uppercase for mappings between entities. Physical objects are given by bold letters, abstract by italic. We now have a picture as in figure \ref{RR}(b). This is the most basic use of representation, and we can immediately see that it is an asymmetric relation. Having an abstract representation for certain physical systems does not, in general, tell us how to find a physical system that matches a given abstract entity. When modelling, the physical system is known to exist (it is that which is modelled). However, there is no a priori reason to suppose that there is a physical system corresponding to every model.
A theorist can write down, for example, the qubit state $\ket{\psi} = \alpha \ket{0} + \beta \ket{1}$; for an experimentalist, however, to discover and build a system to which it corresponds is often no trivial matter. While these two directions of representation are not absolutely disjoint, the exasperation sometimes expressed by experimentalists towards the unrealistic demands of theorists has its roots in the asymmetries of the representation relation between physical and abstract entities.

The two directions of the representation relation, modelling from physical to abstract, and \emph{instantiation} from abstract to physical, lie at the heart of our questions around when a physical system computes. In physics we represent the physical world using abstract and mathematical/logical concepts. In physical computation we want to take an abstract entity, a computation, and represent it physically. Put simply, abstract models may be created at will; physical objects cannot. Without a simple relation that takes us from abstract to physical, how do we use the physical to instantiate the abstract? 

In order to answer this, we first need to consider the interaction between theory and experiment in physics.  To do this, we give a framework in which the relationship between theoretical models and experiments can be understood.  This then forms the basis for a formal framework in which we define physical computation.


\section{Theory and experiment in physics}


The basic purpose of experiments in science is to test a modelling relation: is the model a \emph{good} model? At this stage of testing a theory, the only available representation relation is this modelling relation: we have a theory that takes us from physical to abstract, but not vice-versa. 

\begin{figure}[t]
    \scalebox{0.7}{\hspace{-8cm}

 \begin{minipage}[c]{1.0\linewidth}
  \[
   \begin{array}{ccc}
   \begin{tikzpicture}[font=\large]

\draw[style=dashed] (0,0) -- (10,0);
\draw (0,0.5) node {$\scriptstyle{Abstract}$};
\draw (0,-0.5) node {$\scriptstyle{Physical}$};

\node[draw] (p) at (3,-1) {$\mathbf{p}$};
\node[draw] (mp) at (3,2) {$m_{\mathbf{p}}$};
\draw[->,double] (p) -- (mp);
\draw (3.8,0.5) node {$\RT$};

\end{tikzpicture}

& \qquad  
\quad &   

   \begin{tikzpicture}[font=\large]

\draw[style=dashed] (0,0) -- (10,0);
\draw (0,0.5) node {$\scriptstyle{Abstract}$};
\draw (0,-0.5) node {$\scriptstyle{Physical}$};

\node[draw] (p) at (3,-1) {$\mathbf{p}$};
\node[draw] (mp) at (3,2) {$m_{\mathbf{p}}$};
\draw[->,double] (p) -- (mp);
\draw (3.8,0.5) node {$\RT$};

\node[draw] (mprp) at (6.5,2) {$m^\prime_{\mathbf{p}}$};
\draw[-open triangle 45] (mp) -- (mprp);
\draw (4.8,2.5) node {$C_\theory(m_\mathbf{p})$};

\end{tikzpicture}
   \\{}\\
 (a) & \qquad \quad  &  (b)  
   \\{}\\{}\\
    \begin{tikzpicture}[font=\large]

\draw[style=dashed] (0,0) -- (10,0);
\draw (0,0.5) node {$\scriptstyle{Abstract}$};
\draw (0,-0.5) node {$\scriptstyle{Physical}$};

\node[draw] (p) at (3,-1) {$\mathbf{p}$};
\node[draw] (mp) at (3,2) {$m_{\mathbf{p}}$};
\draw[->,double] (p) -- (mp);
\draw (3.8,0.5) node {$\mathcal{R}_\theory$};

\node[draw] (mprp) at (6.5,2) {$m^\prime_{\mathbf{p}}$};
\draw[-open triangle 45] (mp) -- (mprp);
\draw (4.8,2.5) node {$C_\theory(m_\mathbf{p})$};

\node[draw] (ppr) at (8,-1) {$\mathbf{p}^\prime$};
\draw[-open triangle 45] (p) -- (ppr);
\draw (5.25,-.75) node {$\mathbf{H}(\mathbf{p})$};

\end{tikzpicture}

& \qquad  
\quad &   

   \begin{tikzpicture}[font=\large]

\draw[style=dashed] (0,0) -- (10,0);
\draw (0,0.5) node {$\scriptstyle{Abstract}$};
\draw (0,-0.5) node {$\scriptstyle{Physical}$};

\node[draw] (p) at (3,-1) {$\mathbf{p}$};
\node[draw] (mp) at (3,2) {$m_{\mathbf{p}}$};
\draw[->,double] (p) -- (mp);
\draw (3.8,0.5) node {$\RT$};

\node[draw] (mprp) at (6.5,2) {$m^\prime_{\mathbf{p}}$};
\draw[-open triangle 45] (mp) -- (mprp);
\draw (4.8,2.5) node {$C_\theory(m_\mathbf{p})$};

\node[draw] (ppr) at (8,-1) {$\mathbf{p}^\prime$};
\draw[-open triangle 45] (p) -- (ppr);
\draw (5.25,-.75) node {$\mathbf{H}(\mathbf{p})$};

\node[draw] (mppr) at (8,1) {$m_{\mathbf{p}^\prime}$};
\draw[->,double] (ppr) -- (mppr);
\draw (8.8,-0.3) node {$\RT$};

\end{tikzpicture}
   \\{}\\
(c) & \qquad \quad  &  (d)
 \end{array}
\]
\end{minipage}}
       \caption{Parallel evolution of theory and experiment. (a) Physical system $\p$ is represented abstractly by $m_\p$ using the modelling representation relation $\RT$ of theory $\theory$. (b) Abstract dynamics $\see_\theory (m_\p)$ give the evolved abstract state $m^\prime_\p $. (c) Physical dynamics $\mathbf{H}(\p)$ give the final physical state $\p^\prime$. (d) $\RT$ is used again to represent $\p^\prime$ as $m_{\p^\prime}$.} \setcounter{figure}{2}\label{AL}
\end{figure}
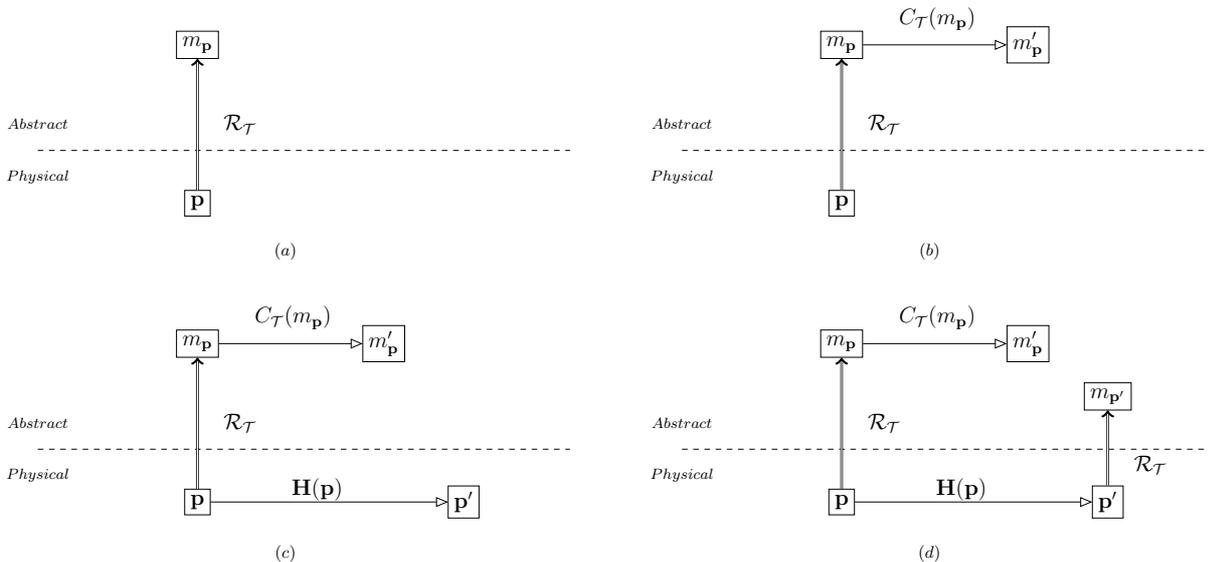

 The models that are used in physics are not isolated, but rather located within specific, abstract, physical theories: an electron has a representation as a wavefunction in standard quantum mechanics, but as a point-mass in classical mechanics and as a vector in Fock space in quantum field theory. This is an important point: the representation relation is theory-dependent. When we test physical theories, we are testing, amongst other things, the representation that they give for physical objects. We therefore write the modelling relation as $\RT$, where $\theory$ is the theory in which it is located.
  
The model of a specific physical system, what we might call the kinematical representation, is then subject to the dynamics of the abstract theory. For example, the wavefunction $\psi$ of an electron in a Stern-Gerlach apparatus would be described as interacting under a given Hamiltonian dependent on the magnetic field strength. This can be worked out purely mathematically. Note that we are using the term `dynamics' somewhat loosely; any theory of the physical system that produces output states from input states is applicable, whether it be couched in terms of evolution over time, or least-action principles etc..

We now have the situation at the abstract level given in figure \ref{AL}(a): a physical system $\p$ is given an abstract representation $m_\p$ by the modelling representation relation $\RT$. This is then evolved using the dynamics of theory $\theory$, $\see_\theory$, resulting in the abstract system $m^\prime_\p$, as shown in figure \ref{AL}(b). Now the physical system $\p$ is not, in general, static: it undergoes its own evolution in the physical world, $\mathbf{H}$. The resultant physical system, after evolution, is $\p^\prime$, as shown in figure \ref{AL}(c). We now have the question about the relationship of $\p^\prime$ to $m^\prime_\p$. $m^\prime_\p$ is the abstract description, probably mathematical, of how the theory $\theory$ thinks our physical system $\p$ should have evolved. How do we tell if $\theory$ has got it right or not? 
To do this, we need some way to compare $\p^\prime$ to $m^\prime_\p$. 
With only a modelling relation we cannot construct a physical system from $m^\prime_\p$ and compare it to $\p^\prime$; however, we can construct a mathematical entity from $\p^\prime$, using $\RT$, and compare it to $m^\prime_\p$.

This gives us the situation in figure \ref{AL}(d): at the abstract level we now have the abstractly-evolved system $m^\prime_\p$ and the abstract representation of the physically-evolved system $m_{\p^\prime}$. Two abstract objects created by the same representation relation $\RT$ can now be directly compared. 

What we expect of a `good' physical theory is that it produces a commuting diagram from this figure. In other words, that the theory $\theory$ is such that we can either let a system undergo physical evolution, or evolve it abstractly, and still reach the same place in the diagram corresponding to the `correct' answer. This not a full specification of what it means to be  a good physical theory, but simply a minimal requirement: that the prediction of the theory, $m^\prime_\p$, is what we get in reality. An absolutely commuting diagram therefore requires that $m^\prime_\p = m_{\p^\prime}$, and it would seem at first sight that this is the requirement given in experimental physics: if the mathematical representation of the experiment outcome is not identical to the prediction, then the theory falls under suspicion. Compare, for example, the diagram used by Ladyman \etal to define their `L-machine' \cite{ladyman}, which uses nondirectional representation and requires absolute commutation. However, this is a much more stringent requirement than is used in practice. Experimental error and limitations of modelling mean that we are content if $m^\prime_\p$ and $m_{\p^\prime}$ are `close enough': $|m^\prime_\p- m_{\p^\prime}| < \epsilon$. Exactly how big or small $\epsilon$ can be to be `good enough' depends very much on the context of the experiment: an undergraduate finding the energy levels of a well-studied SQuID for an assignment will probably impose a less strict closeness requirement than a team testing whether they have found the Higgs boson. The outcome in terms of the diagram, however, is the same: for the practical purposes to which it will be put, for the accuracy at which it has been tested, the theory $\theory$ is such that the diagram commutes. Abstract predictions may then be made of physical evolution, which are the same as the abstract representations of the evolved physical systems.

It is worth emphasising again exactly what is involved in diagrams such as figure \ref{AL}, and those for the Layman L-machine. These are diagrams indicating representation of physical objects (below the line) by abstract ones (above). Physical objects themselves are indicated below the line, not a mathematical representation of them. This contrasts with another set of diagrams that look at first sight very similar: those of Abstract Interpretation, where the concrete (operational) semantics for a computer is related to the abstract semantics for its programming \cite{absint}. While structurally similar to the diagrams here, Abstract Interpretation (as its name suggests) concerns entirely mathematical objects (the concrete and abstract semantics). The relations between them are straightforwardly mathematical relations. The representation relation, however, is not mathematical: therein lies the difference between the treatment of computers in theoretical computer science and our present concern to deal with them explicitly as objects in the physical world.

\section{Commuting diagrams}

We have spoken above somewhat loosely about a theory $\theory$ producing a commuting diagram for experiments. We now detail exactly what $\theory$ consists in, and its relationship to the representation and dynamics, $\RT$ and $\see_\theory$, used in our diagrams. First of all, though, we should note that we have not taken up a stance on what is needed for an experiment to confirm or refute a theory: all we are claiming is that any reasonable description of the scientific process must produce a \emph{de facto} commuting diagram.   

\begin{figure}[t]
    \scalebox{0.7}{

 \begin{minipage}[c]{1.0\linewidth}
  \[
   \begin{tikzpicture}[font=\large]

\draw[style=dashed] (0,0) -- (10,0);
\draw (0,0.5) node {$\scriptstyle{Abstract}$};
\draw (0,-0.5) node {$\scriptstyle{Physical}$};

\draw (3,-1) node {Apparatus};
\draw (3,-1.5) node {+ test system};
\node[draw] (mpxp) at (3,2) {$m_{\p\mathrm{(apparatus+test)}}$};
\draw[->,double] (3,-0.75) -- (mp);
\draw (3.8,0.5) node {$\RT$};

\node[draw] (mprp) at (7.5,2) {$m^\prime_\p$};
\draw[-open triangle 45] (4.7,2) -- (mprp);
\draw (6,2.5) node {$C_\theory(m_\p)$};

\draw (9,-1) node {Apparatus+system};
\draw (9,-1.5) node {final state};
\draw[-open triangle 45] (4.2,-1) -- (7.2,-1);
\draw (5.5,-.75) node {laser fires};

\node[draw] (mppr) at (9,1) {$m_{\p^\prime}$};
\draw[->,double] (9,-0.75) -- (mppr);
\draw (9.8,-0.3) node {$\RT$};

\draw[<->,style=dashed] (mprp) to[bend left=40] (mppr);
   \draw (9,2) node {$\epsilon$};
   
\end{tikzpicture}
\]
\end{minipage}}
\caption{A `good enough' commuting diagram for an experiment to test a theory. See text for details.}\label{exp}
\end{figure}
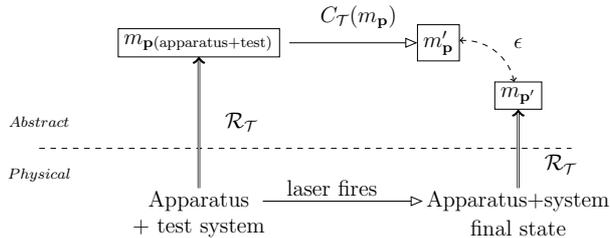

Let us consider an experiment to test a physical theory $\theory_\mathrm{test}$, figure \ref{exp}. The physical setup is denoted by $\p$ as before, and comprises the entire experiment. To take a specific example, consider a rubidium atom in a cavity that is being excited by laser light in order to test a theory of when its excited state will decay for a certain wavelength of incoming photons. $\p$ comprises both the atom that is being investigated and the apparatus (cavity, laser, detection devices etc): $\p= \p_\mathrm{test} + \p_\mathrm{apparatus}$. The apparatus is described by a theory $\theory_\mathrm{apparatus}$. The abstract description of the experimental setup, $m_\p$, is produced using the representation relation corresponding to the theory of the apparatus, $ \mathcal{R}_{\theory(\mathrm{apparatus})}$. 

The experiment then proceeds: the laser is fired, the atom excited, and a decay event timed. The entire physical system evolves to $\p^\prime$, as before. The evolution of the abstract system must now be worked out to find the prediction against which the experimental outcome will be measured. The combination of the theories of the apparatus and the theory being tested produces a set of dynamical equations (or other abstract representation that takes initial to final states). This combined theory, $\theory$, we can write as $\theory = \theory_\mathrm{test} + \theory_\mathrm{apparatus}$. The complete set of dynamics it produces is $C_\theory$. Applying these to the specific system model $m_\p$ to predict its evolution entails calculating the evolution $C_\theory(m_\p)$. The result is the prediction $m^\prime_\p$. 

We now reach the final stage of the experiment. The entire experiment, apparatus plus atom, has evolved to its outcome state. In order to compare with the prediction, an abstract description of this final state is needed. This is produced by another use of the modelling relation for the apparatus, $\mathcal{R}_{\theory(\mathrm{apparatus})}$. This is the step that takes us from, for example, current surges in a detector to a description that a photon was detected at a certain time. We rely on our theory of the experimental apparatus to say that such an observed effect came from a photon, not any other kind of event. The fact that we must make use of $R_{\theory(\mathrm{apparatus})}$ to represent the outcome of experiments is known in the philosophy of science as the \emph{theory-ladenness of observation} \cite{theory}. There are no `basic' observations that are unmediated by any kind of theory, all the way down to the level that when we see, hear, or touch something we must form the theory that our senses are not deceiving us in order to correlate sense data with external objects.

Let us assume that the experiment was a success, and $m_{\p^\prime}$ is close enough (by whatever criteria we are using) to $m^\prime_\p$. The theory that then lives to fight another day is the combined theory $\theory = \theory_\mathrm{test} + \theory_\mathrm{apparatus}$ under the particular circumstances of the experiment which used the dynamics $\see_\theory(m_\p)$ of the combined system $\p= \p_\mathrm{test} + \p_\mathrm{apparatus}$. What has actually been tested in this experiment is this very specific set of dynamics and representation: we have a commuting diagram for $\mathcal{R}_{\theory(\mathrm{apparatus})}$ and $C_\theory(m_\p)$ -- not $\theory$ itself. This is the reason why multiple experiments on many different systems are considered necessary in order to argue for the correctness of a theory $\theory$ (the process by which this actually happens being one of the foundational problems of the philosophy of science that we are not attempting to solve).

Moreover, $\theory = \theory_\mathrm{test} + \theory_\mathrm{apparatus}$, and so if we want to use the experiment to test $\theory_\mathrm{test}$, we need to be sure about $\theory_\mathrm{apparatus}$. This means that $\theory_\mathrm{apparatus}$ must previously itself have been subject to testing by a series of experiments, each of which formed their own commuting diagrams. These will be tests of both the dynamics and the model of the apparatus. If the theory of the apparatus, in either dynamics or modelling, is incorrect, then the experiment is flawed. An example of an incorrect theory of apparatus was the 2011 announcement of faster-than-$c$ neutrino speed by the OPERA experiment \cite{opera1}. A cable connected in an unexpected manner meant that the theory of the apparatus was incorrect, and hence that the representation $\mathcal{R}_{\theory(\mathrm{apparatus})}$ to find the arrival time measurement was flawed \cite{opera2}. This gave an incorrect abstract description $m_{\p^\prime}$ to the experimental outcome (in that specific case, an incorrect time stamp to a detection event). As a consequence, an incorrect argument was made that the failure of $\theory= \theory_\mathrm{relativity} + \theory_\mathrm{apparatus}$ was owing to a failure of $\theory_\mathrm{relativity}$ rather than, as turned out to be the case, a failure of $\theory_\mathrm{apparatus}$.

Experimental science then usually proceeds by using apparatus about whose theory we are reasonably confident to test theories of specific systems about which we are not so confident. As the OPERA result showed, this is in practice usually a messy affair, not a straightforward progression through progressively more `true' theories. An experimentalist whose apparatus does not spring nasty behavioural surprises on them on a regular basis 
is fortunate indeed. We can therefore think of the whole process in terms of multiple inter-connected diagrams, each of which is a specific experimental instance, for different theories (for example, $\theory_\mathrm{apparatus} + \theory_\mathrm{test \ apparatus}$ to test the theory of the apparatus using another apparatus). Whatever the method turns out to be by which scientific theories are chosen (confirmation, refutation, explanatory power \ldots) the desired outcome is all these diagrams commuting. The scientific process can therefore be thought of as solving, by whatever method, this many-diagram satisfiability problem. The outcome of this process is then a set of theories that give rise to commuting diagrams in known cases, which we have confidence (however gained) will also produce commuting diagrams given other specific instances of a physical system $\p$ and its dynamics.

\section{Reversing the modelling relation: prediction and technology}


A theory producing a set of commuting diagrams is not the end of the scientific process. Once armed with a `good' physical theory, it is then put to use (with the proviso, again, that we make no claim about the method by which theories are chosen as `good'). The theory itself can be seen as an explanation of physical phenomena already known (the physical systems $\p$ that were modelled as $m_\p$ and then used in the original experiments). The next step is to use the theory as a \emph{predictive} tool, inferring the existence of phenomena, or even physical objects, about which we were previously ignorant. 

\begin{figure}[t]
    \scalebox{0.75}{\hspace{-6cm}

 \begin{minipage}[c]{1.0\linewidth}
  \[
   \begin{array}{ccccc}
   \begin{tikzpicture}[font=\large]
   
   \draw[style=dashed] (2,0) -- (9,0);

\node[draw] (p) at (3,-1) {$\p$};
\node[draw] (mp) at (3,2) {$m_\p$};
\draw[->,double] (p) -- (mp);
\draw (3.8,0.5) node {$\RT$};

\node[draw] (mprp) at (8,2) {$m^\prime_\p\approx m_{\p^\prime}$};

\draw[-open triangle 45] (mp) -- (mprp);
\draw (4.8,2.5) node {$C_\theory (m_\p)$};

\node[draw] (ppr) at (8,-1) {$\p^\prime$};
\draw[-open triangle 45] (p) -- (ppr);
\draw (5.25,-.75) node {$\mathbf{H}(\p)$};

\draw[->,double] (ppr) -- (mprp);
\draw (8.75,0.5) node {$\RT$};

\end{tikzpicture}
   &   &
      \begin{tikzpicture}[font=\large]
   
   \draw[style=dashed] (2,0) -- (7.5,0);

\node[draw] (p) at (3,-1) {$\p$};
\node[draw] (mp) at (3,2) {$m_\p$};
\draw[->,double] (p) -- (mp);
\draw (3.8,0.5) node {$\RT$};

\node[draw] (mprp) at (6.5,2) {$m^\prime_\p\approx m_{\p^\prime}$};

\draw[-open triangle 45] (mp) -- (mprp);

\end{tikzpicture}
   
      &   &
      \begin{tikzpicture}[font=\large]
   
   \draw[style=dashed] (2,0) -- (9,0);

\node[draw] (p) at (3,-1) {$\p$};
\node[draw] (mp) at (3,2) {$m_\p$};
\draw[->,double] (p) -- (mp);
\draw (3.8,0.5) node {$\RT$};

\node[draw] (mprp) at (7,2) {$m^\prime_\p\approx m_{\p^\prime}$};

\draw[-open triangle 45] (mp) -- (mprp);

\node[draw] (ppr) at (7,-1) {$\p^\prime$};

\draw[->,double] (mprp) -- (ppr);
\draw (7.75,0.5) node {$\IRT$};

\end{tikzpicture}

      \\{}\\
(a) &   &  (b) & & (c)
 \end{array}
\]
\end{minipage}}
       \caption{Reversing the modelling relation within science: (a) A fully commuting diagram for physical and abstract evolution, based on a modelling relation only. (b) The `predict cycle': abstract theory is used to predict physical evolution. (c) The `instantiation cycle': using an instantiation representation relation $\IRT$, a physical object is found corresponding to the predicted evolution. } \label{predict}
\end{figure}
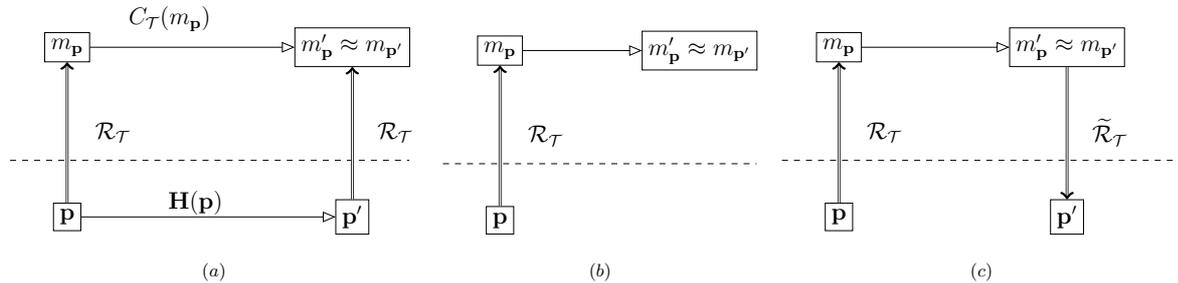

There are two stages to prediction in science. The first is the use of the modelling relation to give an abstract object that is then evolved. Based on a good theory, confidence that the complete diagram, figure \ref{predict}(a) would commute means that the physical evolution is not run: the abstract evolution alone suffices to give the abstract representation of the physically-evolved system. This is the `predict cycle', figure \ref{predict}(b): abstract evolution is used instead of physical to find the result $m_{\p^\prime}\approx m^\prime_\p$.

If what is required out of a theory is an abstract prediction, then the cycle stops here. However, there is a second stage. The abstract theory has now been used to describe an abstract object different from the abstract descriptions of currently known physical objects. To what physical object does the abstract one correspond? Stated in terms of the modelling relation, this question becomes: what physical system, when modelled using our theory, will render this abstract object? In other words, we want to be able to reverse the modelling relation, to find a physical object corresponding to our new abstract description.

Reversing the modelling relation then requires us to have at our disposal an entire set of commuting diagrams, so that we can find the correct one to get a representation that in effect `runs in reverse' from abstract to physical. This is a highly skilled and creative task for both theorists and experimentalists. There are many levels of interlocking diagrams that are involved in developing and testing a theory, and that are then produced when a theory is used to predict outside the range of physical events used to test it. While a reasonable level of confidence in a tested theory is needed in order to predict, prediction-and-instantiation diagrams, figure \ref{predict}(c), also become part of the many-diagram satisfiability problem that is the scientific process, as noted above.

Instances of prediction and subsequent discovery using scientific theories are, of course, numerous. One famous example is Dirac's prediction of positrons \cite{dirac}. By starting with a theory that had been experimentally tested using many physical systems, and using knowledge of the way in which the theory would model situations other that those that had been tested, the prediction was made that a particular abstract object in the theory (a hole in a sea of negative energy electrons) would correspond to a physical object (a positron). This prediction allowed a standard experimental cycle to be set up, and the diagram was found to commute.

The construction and testing of a scientific theory that is robust enough to have predictive as well as explanatory power is the end-point of the scientific process. However, the well-tested commuting diagrams can then be put to use in order not just to discover new physical systems, but to construct them. This is the realm of \emph{technology}: using our theories to precision-engineer physical systems to desired specifications. This is the final element needed in order for us to use this set of commuting diagrams as a framework in which to describe computation.

Engineering and technology are reversals of the modelling relation in a very specific manner. They start from the point of having a well-developed physical theory $\theory$, which we have sufficient confidence in to expect that it will produce diagrams that commute outside the situations in which it was tested. Within the representation of this theory, there is an abstract specification of the physical system that we wish to construct, which we will (leadingly) call $m_{\p^\prime}$. The aim of technology is to construct the corresponding physical system, $\p^\prime$, effectively reversing the modelling relation. 

The process of technology to produce this reversal consists in finding a physical system $\p$, the theory $\theory$, and a specific set of evolutions $\mathbf{H}$ that will perform the evolution $\p \longrightarrow \p^\prime$ such that, when $\p^\prime$ is represented using $\RT$, it becomes the desired $m_{\p^\prime}$. The physical system $\p$ is thus engineered using the process $\mathbf{H}$ to produce the desired physical system $\p^\prime$. An example would be taking a set of steel girders and building a bridge out of them. 

\begin{figure}[t]
    \scalebox{0.75}{\hspace{-5cm}

 \begin{minipage}[c]{1.0\linewidth}
  \[
   \begin{array}{ccccccc}
   \begin{tikzpicture}[font=\large]
   
   \draw[style=dashed] (2,0) -- (5,0);

\node[draw] (p) at (3,-1) {$\p$};
\node[draw] (mp) at (3,2) {$m_\p$};
\draw[->,double] (mp) -- (p);
\draw (3.8,0.5) node {$\IRT$};

\end{tikzpicture}

   & \begin{tikzpicture}[font=\large]
   \draw(0,0) node {};
    \draw(0,1.5) node {$\equiv$};  
    \end{tikzpicture}
     &

      \begin{tikzpicture}[font=\large]
   
   \draw[style=dashed] (2,0) -- (7,0);

\node[draw] (p) at (3,-1) {$\p$};
\node[draw] (ppr) at (6,-1) {$\p^\prime$};
\node[draw] (mppr) at (6,2) {$ m_{\p^\prime}$};

\draw[->,double] (ppr) -- (mppr);
\draw (6.75,0.5) node {$\RT$};

\draw[-open triangle 45] (p) -- (ppr);
\draw (4.5,-0.5) node {$\mathbf{H}$};

\end{tikzpicture}
   
      &  \begin{tikzpicture}[font=\large]
   \draw(0,0) node {};
    \draw(0,1.5) node {\textbf{and}};  
    \end{tikzpicture} &
      
            \begin{tikzpicture}[font=\large]
   
   \draw[style=dashed] (2,0) -- (7,0);

\node[draw] (p) at (3,-1) {$\p$};
\node[draw] (mp) at (3,2) {$m_\p$};
\draw[->,double] (p) -- (mp);
\draw (3.8,0.5) node {$\RT$};

\node[draw] (mprp) at (6.5,2) {$m^\prime_\p$};
\draw (4.9,2.3) node {$C_\theory$};
\draw[-open triangle 45] (mp) -- (mprp);

\end{tikzpicture}
     
      &  \begin{tikzpicture}[font=\large]
   \draw(0,0) node {};
    \draw(0,1.5) node {\textbf{and}};  
    \end{tikzpicture} &
    
          \begin{tikzpicture}[font=\large]
   \draw(0,0) node {};
    \draw(0,1.5) node {$m^\prime_\p\approx m_{\p^\prime}$};  
    \end{tikzpicture} 

 \end{array}
\]
\end{minipage}}
       \caption{Technology reversing the modelling relation: $\p$, $\theory$ and $\mathbf{H}$ are found such that these conditions hold. The combination of these conditions is a commuting diagram. } \label{techno}
\end{figure}
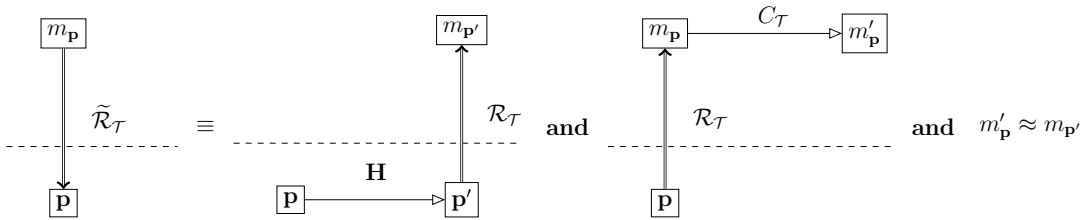

A key consideration is how $\p$, $\theory$, and $\mathbf{H}$ are to be found. With a reliable theory, they can be discovered using abstract tools: in our bridge example, rather than physical trial-and-error of different materials and construction techniques, a given starting point $\p$ can be modelled abstractly as $m_\p$, then evolved to a final abstract state $m^\prime_\p$. If this is close enough to the desired $m_{\p^\prime}$ then the corresponding $\p$ and $\mathbf{H}$ are good candidates for building the system. This is not a mechanical or algorithmic process: the correct $\p$, $\theory$, and $\mathbf{H}$ can be checked (at the very least the bridge can be built and we can see if it falls down), but there is no straightforward process to select ones for testing in the first place. This is an important fact about reversing the modelling relation: it requires ingenuity and skill on the part of the scientists and engineers involved. We can talk about a `reversed modelling relation', or an `instantiation relation', but only with the understanding that this is a shorthand for a whole sequence of preconditions. We write as a shorthand $\IRT$, understanding that $\IRT \equiv f(\RT, \theory)$ relies both on the theory $\theory$ that has been developed, and on the primitive modelling relation $\RT$. The equivalence is given in figure \ref{techno}: the physical system $\p$ evolves under $\mathbf{H}$ to $\p^\prime$ which is represented in $\theory$ as the desired $m_{\p^\prime}$; $\theory$ is such that the representation of $\p$ evolves abstractly to $m^\prime_\p$; and $m^\prime_\p \equiv m_{\p^\prime}$. The conjunction of these three conditions is that the full diagram commutes. We can, then, reverse the modelling relation with technology, but only when the theory $\theory$ is sufficiently advanced confidently to give commuting diagrams in all the cases we wish to consider.


\section{When does a physical system compute?}\label{when}

We are now in a position to demonstrate how computation fits into this framework of physical theory, experiment, prediction, and technology. We argue that a `computer' is a physical system about which we have a set of physical theories from which we derive both the full representation relation $\{\RT,\IRT\}$ and the dynamics $C_\theory$. We are sufficiently confident in our theory $\theory$ that we can assume that it gives rise to commuting diagrams even when the exact starting states, $\p$ and $m_\p$, and the precise evolutions, $\see_\theory(m_\p)$ and $\mathbf{H}(\p)$, are different from the states and evolutions used in testing. Only when our physical theory of the computational device is sufficiently advanced that we can argue that all diagrams commute (in the scenarios we will use it for) can the physical system be used as a computer. In this situation, as when the theory is used for prediction, we must have a sufficiently advanced and good theory that the representation relation can run in either direction.

The first distinction between computing and experimental science in this framework is the initial state. Previously, the physical state $\p$ has been the starting point; however, in a computation the initial impetus is not a physical system that needs to be described, but rather an abstract object that we wish to evolve. An abstract problem is the reason why a physical computer is used. 

We therefore start immediately with the problem of a reversed representation relation. The abstract initial state $m_\p$ must be instantiated in a physical system $\p$: right from the beginning, we see that a computer is fundamentally an item of technology. Even to begin the process of computation, we require a well-understood and well-tested system. The reversal of the representation relation at the start of a computation is the process of \emph{encoding} abstract data in the physical system. It relies fundamentally on knowing exactly how the physical system works; on having a good enough physical theory to predict how data encodings will work. The encoding representation, $\IRT$, is not only dictated by the physics of the computer, but also by our \emph{choice} of how to represent abstract computational objects such as numbers in physical systems. For example, system designers in a standard semi-conductor-based computer \emph{chose} the modelling representation `voltage high $\rightarrow$ 1, voltage low $\rightarrow$ 0'. A crucial part of this choice is to make a modelling relation $\RT$ that is easy to `reverse' to get the $\IRT$ needed for encoding at the initial stage of computation. Another example of an $\RT$ that is easy to `effectively reverse' is the dial input on a Babbage engine \cite{babbage}. An initial $\p$ that is the dial set to a certain angle is then represented as `0', another angle as `1', another as `2', and so on. With appropriate markings on the dial, it is easy for the user to set up an initial physical situation that is represented as the desired number. In contrast, an example of a representation that is extremely difficult to `effectively reverse' is given by the old-fashioned computers that used punch-cards. A pattern of holes on a card determined the input (and indeed the program). Knowing exactly which holes to punch where (\ie the exact physical state $\p$ to produce) such that it had the desired abstract representation (such as `01') was considered extremely tedious and error-prone, requiring a great deal of skill and experience. In more recent times, anyone who has struggled to push the right buttons on their smartphone to do the simplest task has experienced a representation relation that was difficult to reverse, giving a difficult-to-use encoding relation. Making $\RT$ sufficiently easy and intuitive to in-effect reverse is a core component of designing and building a physical computing device. 

%
%

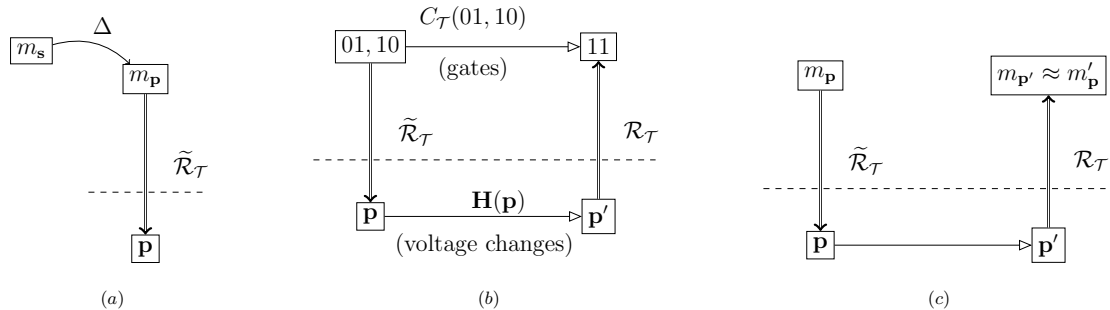
\begin{figure}[t]
    \scalebox{0.75}{\hspace{-5cm}

 \begin{minipage}[c]{1.0\linewidth}
  \[
   \begin{array}{cccccc}
      \begin{tikzpicture}[font=\large]
   
   \draw[style=dashed] (2,0) -- (4,0);

\node[draw] (p) at (3,-1) {$\p$};
\node[draw] (mp) at (3,2) {$m_\p$};
\draw[->,double] (mp) -- (p);
\draw (3.8,0.5) node {$\IRT$};

\node[draw] (ms) at (1,2.5) {$m_\mathbf{s}$};
 
\draw[->] (ms) to[bend left=30] (mp); 
\draw (2.25,2.95) node {$\Delta$}; 
   
   \end{tikzpicture}
     & \qquad \qquad     &
     
   \begin{tikzpicture}[font=\large]
   
   \draw[style=dashed] (2,0) -- (8,0);
   
   \node[draw] (p) at (3,-1) {$\p$};
\node[draw] (mp) at (3,2) {$01,10$};
\draw[->,double] (mp) -- (p);
\draw (3.8,0.5) node {$\IRT$};

\node[draw] (mprp) at (7,2) {$11$};

\draw[-open triangle 45] (mp) -- (mprp);
\draw (4.8,2.5) node {$C_\theory (01,10)$};
\draw (4.8,1.6) node {(gates)};

\node[draw] (ppr) at (7,-1) {$\p^\prime$};
\draw[-open triangle 45] (p) -- (ppr);
\draw (5.25,-.75) node {$\mathbf{H}(\p)$};
\draw (5,-1.5) node {(voltage changes)};

\draw[->,double] (ppr) -- (mprp);
\draw (7.75,0.5) node {$\RT$};

\end{tikzpicture}
     & \qquad \qquad    &
      \begin{tikzpicture}[font=\large]
   
   \draw[style=dashed] (2,0) -- (8,0);

\node[draw] (p) at (3,-1) {$\p$};
\node[draw] (mp) at (3,2) {$m_\p$};
\draw[->,double] (mp) -- (p);
\draw (3.8,0.5) node {$\IRT$};

\node[draw] (mprp) at (7,2) {$ m_{\p^\prime} \approx m^\prime_\p$};

\draw[-open triangle 45] (p) -- (ppr);

\node[draw] (ppr) at (7,-1) {$\p^\prime$};

\draw[->,double] (ppr) -- (mprp);
\draw (7.75,0.5) node {$\RT$};

\end{tikzpicture}

         \\{}\\
(a) & \qquad \qquad  &  (b)& \qquad \qquad  &  (c)
 \end{array}
\]
\end{minipage}}
    \caption{(a) Embedding an abstract problem $M(S)$ into an abstract machine description $m_\p$ using embedding $\Delta$, then encoding into $\p$. (b) Addition of two binary numbers using a computer (see text for details). (c) The `compute cycle': using a reversed representation relation to encode data, physical evolution of the computer is used to predict abstract evolution. Compare with figure \ref{predict}(c).} \label{comp}
\end{figure}

There is one final element to a full computation. This is the process by which an abstract problem (which may not even be posed mathematically) is put into a form such that it can be manipulated by a computer. This is the (abstract) process of \emph{embedding} the abstract problem in the abstract description of the physical system, figure \ref{comp}(a). To take a simple example, imagine you are very bad at mental arithmetic, and are splitting a \pounds 50 restaurant bill equally between six friends using a calculator app. You embed this problem in the decimal division problem ``$50/6$'', and then encode this into the phone by pressing the correct buttons. The embedded problem, $m_\mathbf{s}$, is the reason why we are interested in instantiating the specific set-up of the computational device that is abstractly represented as $m_\p$. For now, we will take the embedding as read and deal just with $m_\p$; embedding will be discussed in more detail below, \S\ref{refine}.

With embedding and encoding relations in place, let us consider as a simple example a digital computer running an algorithm that adds two two-bit numbers, for example $01+10=11$. We first state how each of the individual pieces fit into the diagram of figure \ref{AL}(d), and then show how computation proceeds. 

The elements of the example are given in figure \ref{comp}(b). The abstract initial state, $m_\p = \{01,10\}$ is encoded, through the reversed representation relation (the encoding relation), in the physical system $\p$. This is the step of \emph{initialisation}: $\p$ is the initial state of the computer hardware (voltage across semiconductors, etc.). The representation relation has been derived from the theory we have about the physical components of device, of current and how it changes under voltage changes. Detecting a high voltage corresponds to representing a `1', and low voltage is `0'. The initial physical setup therefore instantiates an initial abstract state. In our example, two parts of the hardware are designated by $\RT$ as `registers', and the voltages in the components of those areas correspond to the representation of the initial state as `01' and `10' (the two numbers we wish to add). 

At the abstract level, the initial state is used as the input to an algorithm: in this example, it is a sequence of gate operations $\see_\theory$ that takes the input `01,10' and adds them. An important part of computation as actually used is that the result of the abstract evolution (here described in terms of gate operations) is not necessarily known prior to the computation. The final abstract state, $m^\prime_\p=(11)$, is not, in fact, evolved abstractly. Instead, at the physical level, a physical evolution $\mathbf{H}(\p)$ is applied to the state, producing the final physical state $\p^\prime$. In our example, this will be the hardware manipulation of voltages. Finally, an application of $\RT$ takes the final physical state and represents it abstractly as some $m_{\p^\prime}$. 

This final use of the representation relation is the \emph{decoding} step: the physical state of the system is decoded as an abstract state. This is frequently simply the encoding step reversed, as in the above examples; however, it need not be. For example, NMR (classical) computing uses a heterogenous representation. For a particular gate, the input bits are encoded as phases and time delays in the radio frequency pulses used to operate the gate, with different choices for each input ``wire''; the output bit is decoded from the value of the observed integrated spectral intensity \cite{Rosell-Merino2010}. Note also, that different decodings can give rise to different computations being performed overall, even when everything else in the system stays the same \cite{decode}.

After the final decoding step, if the computer has the correct answer then $m^\prime_\p = (11)$. If we have confidence in the theory of the computer, then we are confident that $m_{\p^\prime} = m^\prime_\p$, and that this would be the outcome of the abstract evolution.

We can now see what it means to be performing a computation rather than an experiment. As we have seen, in experimental physics a physical system is set up to parallel the abstract situation in order to test the abstract. The upper half of the diagram has been worked out in detail, and we run the lower half to compare with it. Once we have a commuting diagram, however, we no longer need to `run' both halves: as long as the diagram commutes, and as long as our theory allows us to run the representation relation in both directions, we can get from initial state to final state by either abstract working or by physical evolution. Prediction and instantiation took us by an upper route from physical system to physical system via abstract prediction. Computation takes the lower route, starting in the abstract and ending in the abstract, via the physical computer. \emph{This is physical computing: the use of a physical system to predict the outcome of an abstract evolution}. The `compute cycle', figure \ref{comp}(c), is an inverse of prediction and instantiation, in contrast to the latter's use of abstract theory to predict the outcome of physical events.

We can now give the following as a set of necessary 
requirements for a physical system to be capable of being used as a computer.
\begin{itemize}
\item A theory $\theory$ of the physical computational device that has been tested in relevant situations and about which we are confident. 
\item A representation $\{\RT,\IRT\}$ of the physical system that is used for representing the initial state of the physical system (encoding using $\IRT$) and also for the final state, so that output is produced from the computation (decoding using $\RT$).
\item At least one fundamental physical computational operation that takes input states to output states.
\item The theory, representation, and fundamental operation(s) satisfy the relevant sequence of commuting diagrams.
\end{itemize}

All of these elements must be present in order for a physical system to be identified as acting as a computer.

\section{Physical dynamics and computer programs}

Up to now we have considered the dynamics of the computer and the abstract computation as a single, indivisible evolution. We now look more closely at the structure of this evolution, as in general (and particularly in the case of universal computing) it is made up of smaller units. In a standard, digital, computer these are logic gates; other types of computation use units such as relaxation to a ground state (quantum annealing), or other dynamical operations (as in the case of the differential analyser). In the standard, gate-based, case the input is separate from the program, but in other cases the initialisation of the system can contain both the program and the input. In that case, all the work is done by the representation $\RT$, and the theoretical dynamics $C$ and physical evolution $\mathbf{H}$ do not change for different algorithms. The fundamental issues remain the same in both cases and, for the sake of concreteness, we use the example here of a gate-based programmable computer. In this case, the first use of $\RT$ determines initialisation and the input; $C$ is then the abstract program to be run, and $\mathbf{H}$ the physical dynamics that will implement it.

We have referred to $C$ here as both `algorithm' and `program', and we now need to make precise what we mean by this. An algorithm is a very high-level concept, detailing what is to be performed on an input, such as addition. However, in order to actually implement an algorithm, it needs to be broken down into components, and each of these components represented by fundamental operations -- standardly, these are basic gate operations. This is the process of \emph{refinement} and \emph{compilation}. Once the basic operations have been determined for the algorithm, if there is a sequence of operations (as in standard gate-based computers) then they are \emph{composed} to be run on the physical computer.

In the previous section we discussed the embedding of an abstract problem into the physical computer. It is that process that we are now expanding. The embedding relation can be viewed as a composition of many different abstract embeddings, starting with embedding a problem into an algorithm, and then the refinement and composition of the algorithm into machine descriptions that can then be encoded in the physical computer.

\subsection{Refinement}\label{refine}

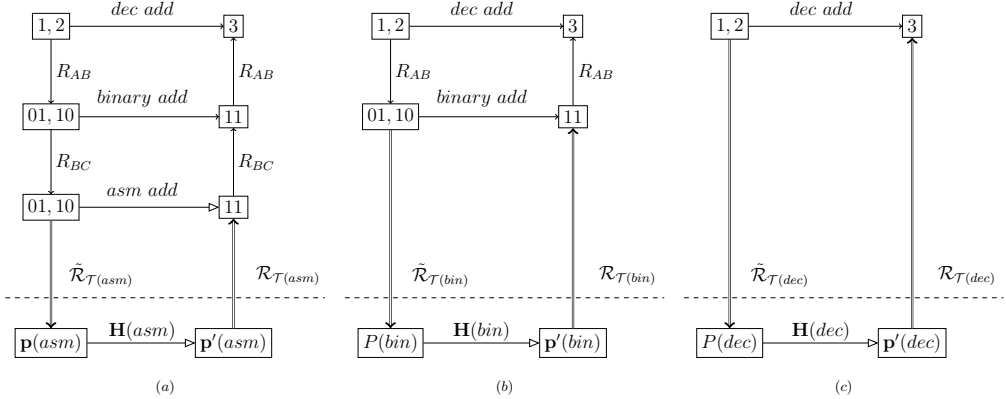
\begin{figure}[t]
    \scalebox{0.6}{\hspace{-6cm}

 \begin{minipage}[c]{1.0\linewidth}
  \[
   \begin{array}{ccccc}
   \begin{tikzpicture}[font=\large]
   
   \draw[style=dashed] (2,0) -- (9,0);
   
   \node[draw] (p) at (3,-1) {$\p(asm)$};
\node[draw] (mp) at (3,2) {$01,10$};
\draw[->,double] (mp) -- (p);

\node[draw] (mprp) at (7,2) {$11$};

\draw[-open triangle 45] (mp) -- (mprp);

\node[draw] (ppr) at (7,-1) {$\p^\prime(asm)$};
\draw[-open triangle 45] (p) -- (ppr);
\draw (5,-.75) node {$\mathbf{H}(asm)$};

\draw[->,double] (ppr) -- (mprp);
\draw (8.2,0.5) node {$\mathcal{R}_{\theory(asm)}$};
\draw (4.1,0.5) node {$\tilde{\mathcal{R}}_{\theory(asm)}$};
\draw (5,2.4) node {$asm \ add$};

\node[draw] (mp1) at (3,4) {$01,10$};
\node[draw] (mprp1) at (7,4) {$11$};

\draw[->] (mp1) -- (mprp1);
\draw[<-] (mp) -- (mp1);
\draw[->] (mprp) -- (mprp1);

\draw (7.5,3) node {$R_{BC}$};
\draw (3.5,3) node {$R_{BC}$};
\draw (5,4.4) node {$binary \ add$};

\node[draw] (mp2) at (3,6) {$1,2$};
\node[draw] (mprp2) at (7,6) {$3$};

\draw[->] (mp2) -- (mprp2);
\draw[<-] (mp1) -- (mp2);
\draw[->] (mprp1) -- (mprp2);

\draw (7.5,5) node {$R_{AB}$};
\draw (3.5,5) node {$R_{AB}$};
\draw (5,6.4) node {$dec \ add$};

\end{tikzpicture}
     &    &
      \begin{tikzpicture}[font=\large]
   \draw[style=dashed] (2,0) -- (9,0);
   
   \node[draw] (p) at (3,-1) {$P(bin)$};

\draw[->,double] (mp1) -- (p);

\node[draw] (ppr) at (7,-1) {$\p^\prime(bin)$};
\draw[-open triangle 45] (p) -- (ppr);
\draw (5,-.75) node {$\mathbf{H}(bin)$};

\draw[->,double] (ppr) -- (mprp1);
\draw (8.2,0.5) node {$\mathcal{R}_{\theory(bin)}$};
\draw (4.1,0.5) node {$\tilde{\mathcal{R}}_{\theory(bin)}$};

\node[draw] (mp1) at (3,4) {$01,10$};
\node[draw] (mprp1) at (7,4) {$11$};

\draw[->] (mp1) -- (mprp1);

\draw (5,4.4) node {$binary \ add$};

\node[draw] (mp2) at (3,6) {$1,2$};
\node[draw] (mprp2) at (7,6) {$3$};

\draw[->] (mp2) -- (mprp2);
\draw[<-] (mp1) -- (mp2);
\draw[->] (mprp1) -- (mprp2);

\draw (7.5,5) node {$R_{AB}$};
\draw (3.5,5) node {$R_{AB}$};
\draw (5,6.4) node {$dec \ add$};
 
\end{tikzpicture}
  &    &
      \begin{tikzpicture}[font=\large]
   
      \draw[style=dashed] (2,0) -- (9,0);
   
   \node[draw] (p) at (3,-1) {$P(dec)$};

\draw[->,double] (mp2) -- (p);

\node[draw] (ppr) at (7,-1) {$\p^\prime(dec)$};
\draw[-open triangle 45] (p) -- (ppr);
\draw (5,-.75) node {$\mathbf{H}(dec)$};

\draw[->,double] (ppr) -- (mprp2);
\draw (8.2,0.5) node {$\mathcal{R}_{\theory(dec)}$};
\draw (4.1,0.5) node {$\tilde{\mathcal{R}}_{\theory(dec)}$};

\node[draw] (mp2) at (3,6) {$1,2$};
\node[draw] (mprp2) at (7,6) {$3$};

\draw[->] (mp2) -- (mprp2);

\draw (5,6.4) node {$dec \ add$};
 
\end{tikzpicture}
         \\{}\\
(a) &     &  (b) &      &  (c)
 \end{array}
\]
\end{minipage}}
\caption{\label{fig:refn} Physical computation, with layers of refinement $R$ on top for base ten (decimal) addition (``$dec \ add$"), binary addition (``$binary \ add$"), and assembly language addition (``$asm \ add$"). Note the physical device and representation differ in each case.}
\end{figure}

Refinement (or reification) is the computational process of taking an abstract algorithm, and producing a suitably equivalent concrete algorithm that is implementable on a computer \cite{refine}.  The requirements for correct refinement (that the concrete design faithfully implements the abstract specification) also involves commuting diagrams; in this case, however, the diagrams live entirely in the abstract realm. As an example, consider the algorithm for decimal addition. Figure~\ref{fig:refn}(a) shows the process of refinement from the abstract concept of mathematical base ten addition, through a more concrete concept of an algorithm for binary addition, to the most concrete (for this example) level of an assembly language program implementation of binary addition.  Each level is in the mathematical realm, and can be proved correct with respect to the higher level.  Some steps (usually the higher level ones) may require human design ingenuity;
lower level steps can be performed automatically (computed) by an interpreter, compiler or assembler.  Refinement of conventional computational algorithms stops in the mathematical realm, and assumes that the underlying physical device correctly implements the lowest level.  The figure shows the standard levels of refinement, positioned on top of our diagram for the underlying device: a physical assembly language computer. The relevant theory is that of the binary arithmetic.  Accompanying theories that need to be developed are those of any relevant compilers and interpreters. Some of these accompanying theories can be purely mathematical (as they ``implement'' formal refinement steps), but some of them have to cross the mathematical-physical divide.

For unconventional computational devices, where the lowest square commutes only ``up to  $\epsilon$'', the traditional refinement approach of sequencing many computations would have to take error propagation into account. 

The dividing line between the physical and mathematical realms is a design choice: more sophisticated physical devices can be engineered to perform appropriate refinement computations.
Figure~\ref{fig:refn}(b) shows the same abstract calculation, here refined only to the level of binary addition,
and being implemented on a physical binary adder. Now the relevant theory is that of the binary addition computer.  This might be a combination of the theories of the assembly language computer and the relevant assembler.

Figure~\ref{fig:refn}(c) shows the same abstract calculation, now with no refinement level, being implemented on a physical arithmetic computer.  Now the relevant theory is that of the physical arithmetic computer.  This might be a combination of the theories of the assembly language computer, the relevant software assembler, and an interpreter or compiler.

These diagrams all assume that the refinement described is possible. This need not be the case: there may be no possible embedding available, at one or more levels. This is the situation in which it is not possible to perform the desired computation on the given hardware. For example, there is no embedding that will allow an arbitrary billion digit integer to be represented in a machine with only a million bytes of memory: the machine simpy isn't large enough. The availability or otherwise of embedding steps tells us about the physical capabilities of our necessarily finite computers, as opposed to the arbitrarily large computations that can be described abstractly.

\subsection{Composition}

The output of a refined and compiled process is a sequence of fundamental abstract operations that compose to produce the desired abstract process (the algorithm). Where a computation is composed of more than one fundamental operation, there are two parts to this: the fundamental operations themselves, and the rules by which they compose. For example, the set of operations could be AND, OR, and NOT, and the composition rules will tell you, for example, what happens when an OR is followed by a NOT. We now look first at what it means to implement one of the fundamental operations in a physical system, and then at their composition where these are now all being run as physical computations.

A gate is an abstract evolution $C_i$. When applied to a particular (abstract) input $x$ it produces the (abstract) output $y=C_i(x)$. It is then the top line of a diagram of its own. To implement this gate physically is to produce a physical system, a representation relation, and a dynamics of the physical system such that the resultant diagram commutes. To do this, the hardware designer uses exactly the same process of theory and experiment that we detailed above as experimental physics: the system is tested with multiple inputs, the representation and the dynamics scrutinised, and finally a theory $\theory_i$ of the gate produced. This theory tells us that when data are represented in such a way in the physical system then the dynamics produces such an output after the final representation. Confidence in this `gate theory' means confidence that whenever the input is given in a specified way, the physical dynamics $\mathbf{H}_i$ that have been chosen by this process of experimentation give rise to a commuting diagram. 

Each individual gate $C_i$ is therefore tested, and the physical theory (which gives the encoding and decoding) $\theory_i$ developed of the gate produces its own commuting diagram with a given physical system $\p_i$, representation $\mathcal{R}_{\theory_i}$ and physical dynamics $\mathbf{H}_i$. 
We also require, as well as individual theories of gates, a theory $\theory_C$ that describes how they compose (making sure that they do not, for example, contradict each other). As with all physical theories, this compositional theory will be produced by the interaction of theory and experiment, and give rise to its own commuting diagrams with the physical system that is being used as a computer.

The theory of the physical computer is therefore developed in order to predict the outcome in situations that are unknown -- exactly as we use theories in physics. This theory is then extended and tested further, in exactly the same way that any physical theory is developed. The end result of this testing and development is a computer, and the theory that governs it, $\theory=\{\theory_C,\theory_i\}$. What confidence in $\theory$ gives is confidence that, within the limits of $\theory$, any diagram that can be written (that is, any input and any program), will commute. The physical system, the computer, can then with confidence be used to find the result of abstract evolutions written as compositions of the fundamental gates. When any (Turing) computable abstract evolution can be so written, the computer is (Turing) \emph{universal} (see for example \cite[ch3]{flc}). A universal computer has the property that the hard work of experimentally producing commuting diagrams need only be done once, then the computer can be used for any computation.


\section{Computational entities}

We now have a set of elements and a framework necessary for identifying when a physical system is performing a computation. Two important parts of this framework are the initial and final steps of encoding and decoding. At the beginning of the computation, the representation relation is used to encode abstract data and programs in the physical system, and then at the end it is used to decode the state of the physical system into an abstract output. Without the encode and decode steps, there is no computation; there is simply a physical system undergoing evolution. This, then, is one of the key ways in which this frameworks distinguishes between a physical system `going about its business', and the same physical system undergoing the same physical evolution, but this time being used to compute. This is how we can escape from falling into the trap of `everything is information' or `the universe is a computer': a system may \emph{potentially} be a computer, but without an encode and a decode step it is just a physical system.

The question of whether a given physical system is acting as a computer then becomes a question of representation at two different levels. Can we represent what is going on, physically and abstractly, as including an encode and decode step, \ie as including representation? A necessary condition of there being representation present is that there is, as well as the computer, an entity capable of establishing a representation relation. That is, an entity that represents \emph{this} specific physical system as \emph{this} specific abstract object, encoding and decoding data into it. Something must always be present that is capable of encoding and decoding: if there is a computer, \emph{what} is using it?

The necessary existence of a computational entity is a fundamental and integral part of the framework presented here. Without this requirement, there is no differentiation between computation and ordinary physical evolution. It also, at first sight, goes completely against the grain of objective science. Perhaps the most important conceptual breakthrough of information science at its inception was the separation of information as a quantity from its meaning \cite{shannoncom}. The former could be discussed independent of any person or thing performing the computation, whereas the latter was irredeemably subjective. If we are now saying that computational processes cannot be described independently of computation entities (human or otherwise), an immediate concern is that the act of computation then becomes wholly subjective, possibly subject to the intent of the entity running the computer, and not something that can be dealt with by an objective scientific theory of computation. This is an important concern, which we now address.

The first thing to note is that all the requirements we have given, including the requirement that a computational entity responsible for representation be present, are objective requirements. It is simply an objective fact of the matter whether or not a computational entity is part of the system. Consider, for example, that you are watching a student work out a problem using a calculator. There is nothing subjective about the existence of the student. Furthermore, the requirements on the computational entity are not subjective (there is no requirement, for example, for an intent to compute or any subjective position to be taken up towards the computational device): the requirement is that an encoding and a decoding are present, an objective fact of the matter. By close observation of the student, you can determine if information is being encoded into and decoded from the calculator. You as the observer can formulate and test the hypothesis that the student and calculator form a computing system. If you and another observer differ in your theories, there is a fact of the matter as to which of you is correct (although, as with any scientific theory, you may not have all the data required to settle the question). Fundamentally, the question of computational entities comes down to the question of the objective existence or otherwise of encoding and decoding. The entities are required only because encoding/decoding cannot be defined otherwise, not because encoding/decoding is subjective or perspectival. 

Computational entities are a requirement for physical computing as opposed to abstract computation. The occurrance of representation is a vital part of physical computing, and computational entities are the ones performing it. This is the central reason that they are required within a computing system: \emph{computational entities are the physical entities that locate the representation relation}. Without representation, encoding and decoding do not happen. Computation considered as a purely abstract process, as in theoretical computer science, does not require a computational entity; however when the abstract is instantiated in a physical computing device, the computational entity responsible for the representation relation between abstract and physical must be physically realised.

There is a very close and important relationship here with another branch of computational theory: communication theory, and how it uses the parties in a transmission to describe the transmission of information. Usually termed Alice and Bob, the communicating entities are responsible for encoding information into a signal at one end, and decoding it at the other. While a theoretical treatment of a communication scenario need deal only with the transmitted signals, actually sending a message requires Alice and Bob. We can, in fact, locate communication entirely within our framework for computing: the encode and decode steps remain (usually performed by distinct spatiotemporally separated entities), and the evolution of the physical system is an identity computation (the message remains the same between sender and receiver). The definitions of computational and communicating entities coincide.

As with a communicating entity, there is nothing in the definition of a \emph{computational} entity that requires it to be human. There is also no need to bring in ill-defined descriptions such as `conscious' or not. Communication theorists refer as a matter of course to computer terminals, or circuits, or photo-detectors as the communicating entities. Simply, anything that is capable of encoding and decoding information is a computational entity. Whether or not any given entity is capable of this is an objective fact of the matter about which hypotheses can be formulated, tested, and argued over. Part of the objective description of the computational entity is the sophistication of the encoding and decoding operation that it is capable of supporting. If the computational entity is a human being, we are fairly certain about what representations it is capable of. If, for example, a person were writing a computer program to solve a second-order differential equation then we would happily describe the encoding and decoding operation as just that. If, on the other hand, a cat walked across the keyboard and randomly touched exactly the right keys to type out that same program, it would not be a good hypothesis that it was calculating a differential equation. To argue that it was would require the cat to be capable of a complexity of encoding and decoding (including a knowledge of differential equations) that we usually describe as outwith a cat's intellectual capacity. This is not something that is subjective or a matter of opinion: it is a matter of fact about which hypotheses can be formed and tested. 

As can be seen from this example, it is also sometimes the case that a degree of argument is needed to settle if something is or is not a computation. Again, this is a situation familiar from communication theory. Take for example the gradual acceptance in the 1960s of the information transmission nature of a bee's ``waggle dance" \cite{von1967dance}. This had not previously been recognised as an instance of communication, and it was only after much debate that a description of the situation as containing an encoding and decoding of information was accepted. This is, however, a matter of fact not of opinion: that argument was required to settle the matter does not make it subjective.

The relationship with communication also illuminates another situation which might otherwise be considered problematic. Entities are required to encode and decode data in the computation; what happens if, say, the computational entity is removed before the decode step? Is computing still happening? The confusion can arise because the physical computer is undergoing the same evolution as during a computation but, in the absense of a decode operation, it is not computing. An example of an exactly equivalent situation with communication helps us see why not. Consider the case of Egyptian hieroglyphics: after the loss of the language, and before the Rosetta stone was deciphered, did a hieroglyphic inscription perform a communication? It was \emph{potentially} a communication, just as a physical system can potentially be a computer. However, until a decode was possible, it did not in actuality perform communication (no-one could read it). Once the language was understood, the decoding relation was in place, and communication could occur. 

Encoding and decoding information in physical objects is something that does not, in itself, restrict computational entities even to being biological. It is perfectly coherent for a computer itself to encode and decode information in another object. For example, we could replace the student in the above situation with a pre-programmed artificial intelligence (AI). While it would probably not be the most efficient use of its processing power, it could certainly use the calculator to find the answer to problems that it did not work out internally. Again, it would be an objective fact of the matter whether it was setting up an encoding and decoding between itself and the computational device, and hence if the physical system (the calculator in this case) were computing.

One final point should be made about computational entities. It is important to be clear exactly what in a computational system is performing the encoding and decoding for the computation. For example, just because a human being is involved in the system does not mean that they are the computational entity. A good example of this is where a human is performing a computational evolution without having access to the encode or decode steps. This was, in fact, the case in the original `computers', which were groups of people performing small repetitive tasks which, when taken as a whole, comprised a computation \cite{hum}. The `computers' were not there the computational entities. A more recent example is the many `crowd-sourcing' games, such as those for circuit optimisation in quantum computers \cite{quant} or gene sequencing in ash trees \cite{ash}. In both cases, human players can become part of the computational evolution without knowledge of the encoding or decoding (in the philosophical literature, this is the position of the inhabitant of Searle's `Chinese room' \cite{searleroom}). As a consequence, they are not computational entities. Instead, the computational entities are the human scientists using the games to compute problems that they have encoded in the games. It would not be impossible, in fact, for an AI-programmed computer to make use of such a game, in which case the AI would be the computational entity, and the human players part of the computer.


\section{Computation and simulation}

We now turn to a specific type of computation, the simulation of the physical dynamics of a system.  While physical computation is a straightforward replacement of physical evolution for abstract computation, it can cause confusion when a physical system is the subject of a computation, as well as a physical system being used to perform a computation. 

\begin{figure}[t]
    \scalebox{0.6}{\hspace{-2cm}

 \begin{minipage}[c]{1.0\linewidth}
  \[
   \begin{array}{ccc}
    \begin{tikzpicture}[font=\large]
   
   \draw[style=dashed] (0,0) -- (6,0);  
   
   \node[draw] (s) at (2,-1) {$\mathbf{s}$};
   \node[draw] (p) at (1,-2) {$\p$};
   \node[draw] (spr) at (4,-1) {$\mathbf{s}^\prime$};
   \node[draw] (ppr) at (5,-2) {$\p^\prime$};
   
   \draw[-open triangle 45] (s) -- (spr);
   \draw[-open triangle 45] (p) -- (ppr);   
      \end{tikzpicture}
   & \qquad \qquad  &
   \begin{tikzpicture}[font=\large]
   
   \draw[style=dashed] (0,0) -- (10,0);
   
   \node[draw] (s) at (3,-1) {$\mathbf{s}$};
   \node[draw] (p) at (1,-2) {$\p$};
   \node[draw] (spr) at (5,-1) {$\mathbf{s}^\prime$};
   \node[draw] (ppr) at (8,-2) {$\p^\prime$};
   
   \node[draw] (ns) at (3,1) {$n_\mathbf{s}$};
   \node[draw] (mp) at (1,2) {$m_\p$};
   \node[draw] (nspr) at (5,1) {$n_{\mathbf{s}^\prime}$};
   \node[draw] (mppr) at (6.3,2.5) {$m^\prime_\p$};   
   \node[draw] (mprp) at (8,2) {$m_{\p^\prime}$};
   
   \draw[-open triangle 45] (s) -- (spr);
   \draw[-open triangle 45] (p) -- (ppr);
   \draw[->,double] (s) -- (ns);
   \draw[->,double] (p) -- (mp);
   \draw[->,double] (spr) -- (nspr);
   \draw[->,double] (ppr) -- (mprp);
   
   \draw[->] (mp) to[bend left=30] (ns);
   \draw[<-] (mppr) to[bend right=30] (nspr);
   \draw[<->,style=dashed] (mppr) to[bend left=40] (mprp);
   
   \draw (0.1,-0.5) node {$\mathcal{R}_{\theory(\p)}$};
   \draw (8.9,-0.5) node {$\mathcal{R}_{\theory(\p)}$};
   \draw (2.25,-0.5) node {$\mathcal{R}_{\theory(\mathbf{s})}$};
   \draw (6,-0.5) node {$\mathcal{R}_{\theory(\mathbf{s})}$};
   
   \draw (7.5,3) node {$\epsilon$};
   \draw (5,2) node {D};
   \draw (2.5,2) node {E};
   
   \end{tikzpicture}
         \\{}\\
(a) & \qquad \quad \qquad \quad  &  (b)
 \end{array}
\]
\end{minipage}}
       \caption{(a) Two separate physical systems, $\p$ and $\mathbf{s}$. (b) Commuting diagram when testing the ability of $\mathbf{s}$ to simulate $\p$.} \label{sim}
\end{figure}
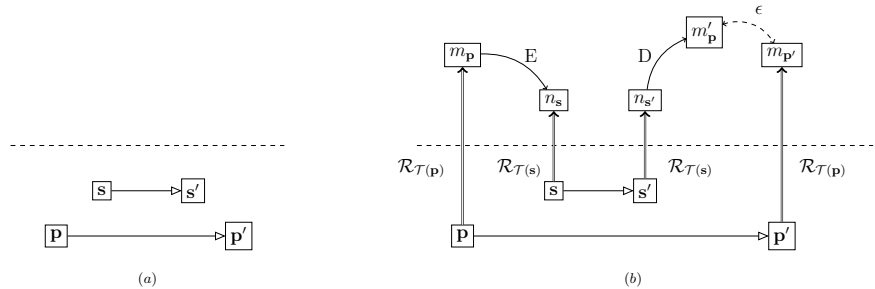

We will consider the situation where a physical system (the computer), $\mathbf{s}$, is to be used to simulate the behaviour of another physical system, $\p$. We show this as in figure \ref{sim}(a). The aim is to build a commuting diagram similar to those given above for standard computation, where the dynamics of $\mathbf{s}$ is used instead of the dynamics of $\p$. 

We are, in fact, in an exactly analogous situation to the introduction of the representation relation: what we want is for system $\mathbf{s}$ to \emph{represent} system $\p$. There is no way of comparing two physical objects without forming a representation of them -- even basic, apparently representation-free, comparisons such as `hold side by side and see if they're the same dimensions' in fact require us to represent parts of the external world by identifying individual objects and a set of properties that are its dimensions (this is a foundational issue in science and metaphysics -- see for example \cite[ch1,2]{vanF}). Abstract representations of the physical systems are created and then used in order to compare the two systems. 

Just as when we looked at diagrams for computers, we start first with the diagram for setting up a simulator and testing that it indeed does what we want. This is given in figure \ref{sim}(b). The steps in the diagram marked `E' and `D' are embedding steps: we wish to embed the abstract description of system $\p$ in the abstract description of system $\mathbf{s}$. For example, if $\mathbf{s}$ is a scale model of $\p$ then this embedding is the relevant scale factor. We saw the process of embedding first with straightforward computation, where an abstract problem is embedded in the abstract description of the physical computer. Here, the abstract problem is itself the abstract description of a second physical process: that is the problem that the computer is being used to solve. As with the refinement process, performing the embedding may itself require computation, in this example, multiplying by the scale factor. In this way, the abstract object $n_{\mathbf{s}}$ is used to represent the abstract object $m_\p$, analogous to an abstract object being used to represent a physical one. 

For system $\mathbf{s}$ to be a good simulator of system $\p$, all relevant diagrams of the form of figure \ref{sim}(b) must commute, closing the gap at the end between $m_{\p^\prime}$ and $m^\prime_\p$. This is discovered in the same way that commuting diagrams for standard computation are found: a sufficiently good theory of the devices is needed, such that we are confident that all diagrams will commute and that the representation can be run in either direction. With this in place we can then change from \emph{testing} the simulator to \emph{using} it. 

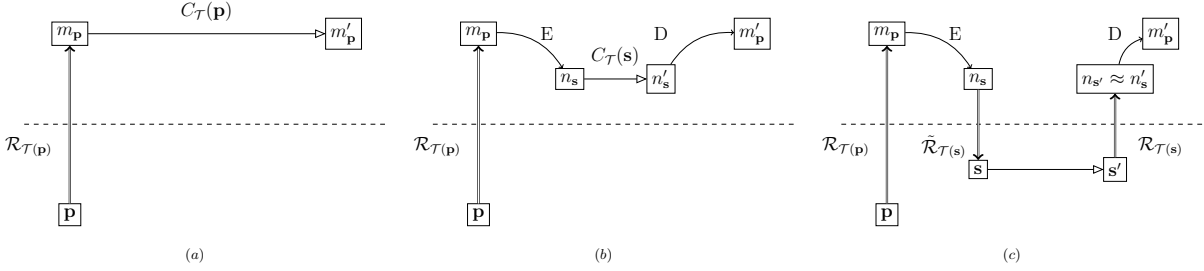
\begin{figure}[t]
    \scalebox{0.6}{\hspace{-12cm}

 \begin{minipage}[c]{1.0\linewidth}
  \[
   \begin{array}{ccccc}
          \begin{tikzpicture}[font=\large]  
       \draw[style=dashed] (0,0) -- (8,0);
  
   \node[draw] (p) at (1,-2) {$\p$};
   
   \node[draw] (mp) at (1,2) {$m_\p$};
   \node[draw] (mppr) at (7,2) {$ m^\prime_\p$};

   \draw[->,double] (p) -- (mp);
   \draw[-open triangle 45] (mp) -- (mppr);
   
   \draw (0.1,-0.5) node {$\mathcal{R}_{\theory(\p)}$};
   \draw (4,2.5) node {$C_\theory(\p)$};
    
          \end{tikzpicture}
   &   &
       \begin{tikzpicture}[font=\large]  
       \draw[style=dashed] (0,0) -- (8,0);
  
   \node[draw] (p) at (1,-2) {$\p$};
   
   \node[draw] (ns) at (3,1) {$n_\mathbf{s}$};
   \node[draw] (mp) at (1,2) {$m_\p$};
   \node[draw] (nspr) at (5,1) {$n^\prime_\mathbf{s}$};
   \node[draw] (mppr) at (7,2) {$ m^\prime_\p$};
   
   \draw[-open triangle 45] (ns) -- (nspr);
   \draw[->,double] (p) -- (mp);
   
   \draw[->] (mp) to[bend left=30] (ns);
   \draw[->] (nspr) to[bend left=30] (mppr);
   
   \draw (4,1.5) node {$C_\theory (\mathbf{s})$};
   \draw (0.1,-0.5) node {$\mathcal{R}_{\theory(\p)}$};
   
   \draw (5,2) node {D};
   \draw (2.5,2) node {E};
    
          \end{tikzpicture}
   &    &
    \begin{tikzpicture}[font=\large]  
       \draw[style=dashed] (0,0) -- (8,0);
  
   \node[draw] (s) at (3,-1) {$\mathbf{s}$};
   \node[draw] (p) at (1,-2) {$\p$};
   \node[draw] (spr) at (6,-1) {$\mathbf{s}^\prime$};
   
   \node[draw] (ns) at (3,1) {$n_\mathbf{s}$};
   \node[draw] (mp) at (1,2) {$m_\p$};
   \node[draw] (nspr) at (6,1) {$n_{\mathbf{s}^\prime}\approx n^\prime_\mathbf{s}$};
   \node[draw] (mppr) at (7,2) {$ m^\prime_\p$};
   
   \draw[-open triangle 45] (s) -- (spr);
   \draw[->,double] (ns) -- (s);
   \draw[->,double] (p) -- (mp);
   \draw[->,double] (spr) -- (nspr);
   
   \draw[->] (mp) to[bend left=30] (ns);
   \draw[->] (nspr) to[bend left=30] (mppr);
   
   \draw (0.1,-0.5) node {$\mathcal{R}_{\theory(\p)}$};
   \draw (2.25,-0.5) node {$\tilde{\mathcal{R}}_{\theory(\mathbf{s})}$};
   \draw (7,-0.5) node {$\mathcal{R}_{\theory(\mathbf{s})}$};
   
   \draw (6,2) node {D};
   \draw (2.5,2) node {E};
    
          \end{tikzpicture}
           \\{}\\
(a) &   &  (b)&   &  (c)
 \end{array}
\]
\end{minipage}

    }
   \caption{System $\mathbf{s}$ running as a simulator for system $\p$, constructed as a compute cycle nested within a predict cycle: (a) the predict cycle using $C(\p)$ to find the abstract prediction for the evolution of system $\p$; (b) embedding $C(\p)$ in the simulator model dynamics, $C(\mathbf{s})$; (c) adding the compute cycle: the physical simulator $\mathbf{s}$ now determines the abstract evolution $C(\mathbf{s})$.}\label{fullsim}
\end{figure}

Full use of system $\mathbf{s}$ to simulate system $\p$ is shown in figure \ref{fullsim}(c). The aim is to reach the abstract outcome $m^\prime_\p \approx m_{\p^\prime}$ without going through the physical evolution $\p \rightarrow \p^\prime$. Instead, three levels of representation are used to achieve the result using the physical system $\mathbf{s}$:
\begin{enumerate}
\item The physical system to be simulated, $\p$, is represented abstractly, $m_\p$.
\item $m_\p$ is embedded into an abstract initial state of the physical system $\mathbf{s}$, $n_\mathbf{s}$.
\item The abstract description $n_\mathbf{s}$ is instantiated as a physical initial state of the simulator, $\mathbf{s}$.
\end{enumerate}

At the end, the state of $\mathbf{s}^\prime$ is decoded to find the output of the simulator, $n_{\mathbf{s}^\prime}$. This is then de-embedded to represent an output state of system $\p$, $m^\prime_\p$. Overall, the abstract description of the simulator is used to represent the abstract description of the system to be simulated, and then the physical simulator device is used to represent its own abstract description.

Figure \ref{fullsim} shows a point that is key to understanding simulation: \emph{what is simulated is the model of the physical system $m_\p$, not $\p$ itself}. The simulator and the physical system under simulation interact only at the abstract level.

There are several, qualitatively different, ways in which simulation is used, which we can show within this framework. Consider the decomposition of a simulation shown in figure \ref{fullsim}. Simulation is viewed as form of prediction, by comparison with figure \ref{predict}(b): the aim is to find the outcome of the physical evolution of $\p$ without actually evolving $\p$ to $\p^\prime$. However, rather than the prediction being performed purely abstractly, the abstract evolution is worked out using a computation. Simulation of one system by another is therefore a compute cycle nested within a predict cycle. Importantly, the physical evolution $\p^\prime$ is taken to match up abstractly with the computational evolution, $m^\prime_\p \approx m_{\p^\prime}$. This is the case when the simulation is accepted as a good guide to the physical evolution; examples include when novel hardware is simulated for the purposes of testing and programming before it is built. 

An alternative way of viewing this figure is to consider the dynamics of the simulator, $C(\mathbf{s})$, and the dynamics being simulated, $C(\p)$. The compute cycle is on this view the fundamental part, the computer being used to determine the dynamics $C(\p)$; this is the computation that is being run. It just so happens that $C(\p)$ is an abstract representation of another physical system. These dynamics are then embedded into the abstract dynamics of the simulator, $C(\mathbf{s})$ -- for example, into an algorithm. This is then run as a compute cycle, with the relevant encoding and decoding into the physical computer. If the final diagram is known to commute, then this is an equivalent description to those given above. However, there is also a case where this form of simulation is used but we do not know whether in the end $m^\prime_\p \approx m_{\p^\prime}$. This is the case of computational physics, where computers are used to simulate behaviour in a physical system during an experiment. Comparing with figure 3, in this situation the abstract dynamics $C_\theory(m_\p)$ are embedded in the simulator and the outcome of this theoretical model is computed using a physical computer. The decoded and de-embedded result of simulating the model is then compared with the abstract description of the experimental outcome. This whole situation is then a compute cycle nested within an experiment cycle.

This framework for simulation is not restricted to the case where the simulating system $\mathbf{s}$ is a standard computer, such as a supercomputer being used to simulate molecular properties of materials. There are other ways in which simulation of a system can be run, where something that is not usually considered to be `a computer' can simulate another physical system. Aircraft designers use wind tunnels and models to simulate the effect of flying on aeroplane parts. A pendulum can be used to simulate a spring and discover oscillation periods. Single-purpose physical simulators have a long history prior to the widespread use of programmable computers, and all of these fit within the framework we have given. 

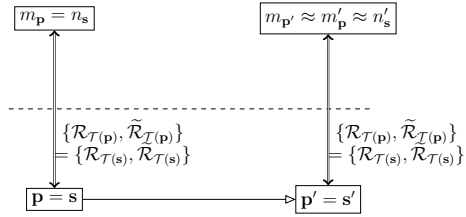
\begin{figure}[t]
    \scalebox{0.6}{

 \begin{minipage}[c]{1.0\linewidth}
  \[
          \begin{tikzpicture}[font=\large]  
       \draw[style=dashed] (0,0) -- (8,0);
  
   \node[draw] (p) at (1,-2) {$\p=\mathbf{s}$};
   \node[draw] (ppr) at (7,-2) {$\p^\prime=\mathbf{s}^\prime$};
   
   \node[draw] (mp) at (1,2) {$m_\p = n_\mathbf{s}$};
   \node[draw] (mppr) at (7,2) {$ m_{\p^\prime} \approx m^\prime_\p \approx n^\prime_\mathbf{s}$};

   \draw[<->,double] (p) -- (mp);
   \draw[<->,double] (ppr) -- (mppr);
   \draw[-open triangle 45] (p) -- (ppr);
   
   \draw (2.5,-0.5) node {$\{\mathcal{R}_{\theory(\p)},\mathcal{\widetilde{R}}_{\theory(\p)}\}$};
   \draw (2.5, -1) node {$=\{\mathcal{R}_{\theory(\mathbf{s})},\mathcal{\widetilde{R}}_{\theory(\mathbf{s})}\}$};
   \draw (8.5,-0.5) node {$\{\mathcal{R}_{\theory(\p)},\mathcal{\widetilde{R}}_{\theory(\p)}\}$};
   \draw (8.5, -1) node {$=\{\mathcal{R}_{\theory(\mathbf{s})},\mathcal{\widetilde{R}}_{\theory(\mathbf{s})}\}$};

          \end{tikzpicture}
\]
\end{minipage}

    }
   \caption{A system simulating itself (compare with figure 8(c)).}\label{selfsim}
\end{figure}

When considering these `non-standard' simulators, one situation that must be addressed is when a simulator is simulating itself. It is often given out as a truism that ``everything simulates itself''. 
It should be clear by now that this is not the case within our framework: just as not every physical evolution is a computation, not every physical event is a simulation. Just as with computation, in the absence of embedding, and of encoding and decoding operations, simulation is not occurring. It is important to note that, in the case of simulation, there are \emph{three} embedding/encoding operations that must be identified. Firstly, the system being simulated must have an abstract representation. Secondly, that abstract representation must be embedded in the abstract representation of the simulator. The final encoding and decoding, into and out of the physical simulator device, is the same as for computation. Without all these steps being present, there is no simulation.

We can now consider the case of a system being used to simulate itself. For example, a pendulum can simulate the same pendulum in a different gravitational field, or a laptop can simulate itself through a virtual environment. In these cases, the places marked `E' and `D' in figure \ref{fullsim}(a) do the work: the embedding is scaling, or virtual software, and so on, even when $\mathbf{s}$ and $\p$ become the same physical system. Note also, that the representations used for $\mathbf{s}$ and $\p$ need not be identical, even when the physical systems are.


Finally, we can push this all the way and consider a situation where not only are the physical systems identical, but so are the representations, and also the embeddings at `E' and `D' are the identity. Do we then have a description by which any physical system is self-simulating? If we look at the resulting diagram, figure \ref{selfsim}, the answer is clearly `no'. We have either a compute cycle, if the abstract theory is well enough known, or part of an experiment. In either case, we still have initial and final representations. In the absence of these representational stages, a system does not simulate itself.

\section{Non-standard computing: computation or experiment?}

We now turn to our main motivation for developing this framework for computing: the analysis of physical devices to see if they are being used as computers. 
As noted previously, the use of a physical system as a computer is first and foremost a use of \emph{technology}: computers are highly engineered devices. We have considered science both in its process of experimentally testing theories, and its ability to predict based on commuting abstract/physical representation diagrams. 
\emph{Engineering} is different in the following way. Consider figure 2(d), in the case of an unacceptably large $\epsilon$.
In science, we have some given $\p$, and are attempting find a good abstract characterisation $C$. If $\epsilon$ is too large, we need to change $C$: we need to find a better characterisation.
In engineering, we have a given $C$ that we wish to physically instantiate, and the
goal is to find $\p$, given $C$. If $\epsilon$ is too large, we need to change $\p$: we need to re-engineer our candidate system.

The essential difference then is the degree of confidence we have in our physical theory of the device: if $\epsilon$ is too big in an experiment then the theory may be disproven; by contrast, in engineering it is the system that is taken to be at fault. Of course, there are frequent cases where this is not a clear-cut distinction. An example is in the earthquake proofing of buildings, where the technology is built according to theory, but cannot be tested at scale. Data collected from each actual earthquake are then used to refine the theory, which is then used in the next generation of technological construction. In general, though, technology stands or falls on the confidence in the underlying theory. Amongst other things, this confidence is that the theory works outside the situations in which it has been tested (note that \emph{any} subsequent use of a theory after it has been tested is a use outside the testing situation: at the very least it differs in time). 

The use of the theory outside the domain in which it has been tested is fundamental to computing: this is \emph{prediction}. As we saw in \S\ref{when}, physical computing is in a sense the inversion of mathematical science, using a physical system to predict the outcome of an abstract dynamics (rather than an abstract model predicting physical dynamics). Without this predictive element, a physical system is not a computer, in the same way that a set of mathematical equations is a bad physical model if it has no predictive power. 



A common, and unfortunate, method of ascribing computational ability to a non-standard system is as follows. A novel computing substrate is proposed (a stone, a soap bubble, a large interacting condensed-matter system, etc., etc.). The physical substrate is `set going' and an evolution occurs. At a certain point the end of the process is declared and measurements taken. The initial and final states of the system are compared, and then a computation and a representation picked such that if the initial state and final states are represented in such a way then such a computation would abstractly connect them. The system is then declared to have performed such a computation: the stone has evalutated the gravitational constant, the soap bubble has solved a complex optimization problem, and so on.

Such arguments, without any further testing or evidence, are rightly treated with suspicion. Given the vast range of representation for physical systems available, almost any computation can be made to fit the difference of initial and final physical states of a system. If such arguments really were correct, we would not only have to conclude that everything in the universe computes, but that everything computes every possible computation all of the time. Such extreme pancomputationalism is even less useful than the usual kind. 

Our framework enables us to see why such arguments are not valid. If a computational description of a physical evolution can only be applied post-hoc, then the system has not acted as a computer. Such descriptions may be used in the experiment and testing cycles for developing a system to use as a computer, but if the final state of the system is needed in order to decide which calculation it has run, the system is not being used to \emph{predict} anything. In such situations the outcome of the abstract computation needs to be known in advance in order to fit the computation to the physical evolution: the physical evolution cannot then be used to give any further data. A post-hoc-only description of computation also fails to predict as the representation needs to be adjusted in order to fit to the computation, and different representations are frequently needed for each `instance of computation'. For a true computer, a general representation for encoding and decoding of data is needed (that doesn't require post-hoc adjustment to make the computation work), and a relevant degree of predictability. A computer is used to predict; the challenge then for non-standard computation is to demonstrate that the theory of the device, and the representation of data within it, is known and stable enough to use the physical device to predict the desired abstract computation.

Classical digital computers are highly-engineered silicon devices with an extremely well developed physical theory in which we have a great deal of confidence. We are confident that we know what they are doing during a computation, and can also predict how they will act in situations outside the usual range. Scaling the system is a matter of correct composition of gates, about which we also have a well developed and good theory. 
The digital nature of the computers is particularly useful, allowing 
systems on which long computations can run to be designed without having to cope with accumulation of smaller errors.
Despite all this confidence, hardware bugs do still occur, one well-known example being the Intel Pentium floating point unit bug \cite{Halfhill1995wr}. Technically, this was caused by a software bug that was then frozen into the hardware design; the boundary between software and hardware is not sharp. Another example is that of modern multi-core implementations which can exhibit unexpected behaviours: the computational abstractions have not developed in step with the physical implementations \cite{Sarkar2011rq}. 


In contrast to the highly-developed and scalable theories of classical computers, non-standard computing devices generally have a theory that is much less well developed. This leads to problems of scale, composition, and confidence that cast doubt on the use of a system as a computer. Amongst unconventional paradigms, quantum computing has the best characterised physical theory.
As with classical digital computers, quantum computers are highly engineered, with quantum states used to represent ``qubits'', the smallest unit of quantum information \cite{blueprint,Ladd2010}. 
Despite the excellent physical theory of quantum mechanics available, however, there is still argument over whether certain specific systems are truly implementing quantum computation. 
Consider, for example, the D-Wave machines \cite{dwaveweb,Johnson2011, dwave100}. Originally presented as implementing a relatively simple quantum annealing paradigm, the consensus has shifted (not least within D-Wave itself) that this physical theory is not a good fit for predicting the computational abilities of the machines. Work is now underway to characterise the devices at a mathematical and phenomenological level, treating them as black boxes \cite{physicsdwave,dwaveeth}. 

This characterisation process crops up frequently in unconventional computing: rather than describing all the physics, as with classical computing, outputs are matched to inputs mathematically. This phenomenological theory can have predictive power (this input is taken to that output); however, the fact that it is only phenomenological directly impacts on the degree of confidence with which the theory is held. Without an underlying physical theory, a phenomenological theory has to do a lot more work to convince that all relevant changes have been taken into account and that the computation can be relied on. Furthermore, different sized systems must be characterised separately, as there is no scalable underlying theory of the device. Without a reliable theory, in what way can the physical evolution of the device predict the abstract evolution of the computation that is supposed to be being run?

 In the end, the question of whether or not an unconventional system can be used as a computer comes down to a simple question: what is the confidence that the abstract/physical diagram for the computation commutes? Without this confidence, it is not computing. There are two options for what is happening, based on the consequence of a mismatch between theory and physical system (\ie a large $\epsilon$): if the conclusion is that the system must be re-designed then the system is being engineered; if the theory of the fundamental dynamics is taken as at fault then what is happening is an experiment. Either way, it is not computing.

Phenomenological models are widely used to develop new physical substrates for unconventional computation, often by adapting devices originally designed for other purposes.  In \cite{Harding2006} a liquid crystal display (LCD)
was configured as a computational device not through engineering design, but through the use of an evolutionary search algorithm to determine the correct configuration. As a consequence, the physical model of the substrate is simply unknown.
As before, one may develop a descriptive physical model within the experimental domain, and exploit that model to compute within the domain, possibly making use of continuity arguments. However, we cannot meaningfully compute with the device outside that domain, since we have no means to extrapolate the model: the descriptive nature of the physical model means the LCD device cannot be scaled with any confidence.

Another example of a purely descriptive physical model is slime moulds. These have, famously, been used to compute minimal path lengths and other geometrical properties \cite{Adamatzky2010}, but with no firm understanding of the underlying physics/biology/chemistry. It is worth noting here that these examples, of slime moulds and evolved LCD computing, help
demonstrate that the physical computer does not need to be intelligently designed: it can be naturally (or even computationally) evolved.
Hence living organisms of all sorts can potentially perform information processing, and can potentially be exploited to perform their computations for us. 

Another major problem arising from the use of such substrates for unconventional computation is that of scaling.  As we have seen, digital computers scale simply through composition of small elements; slime moulds and LCD devices, however, scale by using bigger versions of the same system (this is also arguably true of the D-Wave machine). With only a phenomenological model of performance, there is no guarantee that a scaled-up system will act in the desired way. Even if some scaling behaviour is found experimentally at smaller scales, it is notoriously difficult to project this to larger sizes.

Not all non-standard computing relies on phenomenological models. Sometimes unconventional devices do have a physical model behind them, but this is not in fact the actual physical behaviour (especially at large scales).
Soap films are often described as finding a minimum energy, minimum length, state and thereby performing
an analogue computation of minimal Steiner trees.
However, soap films do not always find minimal states \cite{Aaronson2005}:
it is, after all, the principle of \emph{stationary} action, not \emph{minimal} action, and soap films, as with other physical systems,
can and do get trapped in local minima.
In this case we have a bug in the physical implementation, not because it has been incorrectly engineered, but because the underlying simplistic physical model is wrong.


One final note is that, even for unconventional substrates, the computational model $\C$ is often that of classical boolean logic.  There are only three computational models that claim universality: classical Turing machines, quantum Turing machines \cite{Deutsch1985}, and the general purpose analogue computer \cite{Shannon1941}. 
Although computation does not \emph{require} universality, one interesting area of future research is to develop novel computational models that can be implemented by engineered unconventional substrates. Until then, we can use the framework developed here to distinguish when we are computing with novel substrates and with what degree of confidence, and when we are performing experiments on them.

\section{Conclusion}

We have developed a formal framework for computing, showing how the physical and abstract levels of a computation connect through the representation relation. This relation is the same as governs the interrelation of physical systems and their mathematical description in experimental science, and is key to the scientific process. We have seen how physical science progresses through the experimental realisation of diagrams that commute across the abstract/physical divide, allowing abstract theory to predict the outcome of physical experiments. These diagrams then form the basis for our framework for computation, where the physical evolution of a computer is used to find the outcome of an abstract computation. We are able to use this framework to give conditions on when a physical system is performing a computation: we require a good physical theory of the computer; representation that allows the encoding and decoding of information; and at least one fundamental dynamical operation (such as gates).  The requirement for encoding and decoding to be present in the system leads to the requirement that computational entities be present. Their job is physically to locate the representation relation, which is required for physical computing rather than abstract computation. We saw that the requirement for computational entities does not impact on the objective nature of the conditions for computing. The definition is also broad, including biological, non-biological 
and artificial entities.

The range of potential applications of this framework is huge. Previously, discussion of whether a system is a computer or not has been marked by a large amount of confusion and a correspondingly small amount of consensus. There has simply not previously been the language in which to frame these questions adequately, and to pick apart what is being discussed as an abstract computation, and what as a physical computer. The framework we have presented, including its powerful diagrams, allows us now to define precisely what is being asked. We have seen in this paper some of the first results from this new expressive capability. Physical computing is seen as interacting at a very basic level with experimental science; and we are able to show precisely, within the same framework, the processes of theoretical and experimental science, computation, and engineering and the use of technology. One particular area at the interface between science and computing is the simulation of physical systems by computers. This is an area that often causes a large amount of confusion regarding what is being simulated and by what, and how the simulation and the physical system are related. By locating simulation within our framework, we showed straightforwardly how it relates to computation and theoretical predication, and also how different types of simulation relate to each other. The interaction of different layers through representation was again key: as well as the encoding/decoding operations needed for computing to be occurring, simulation also requires the embedding of abstract models to be taking place. Without these stages, simulation (even of a system of itself) does not occur. Bringing this clarity to such a previously confusing area demonstrates the power of our approach.

Another area that this framework clarifies is that of unconventional computing devices. By considering the theory of the physical device and its limits, we showed that there is a strong danger of misunderstanding what is taking place when these devices are used. It is often the case that the systems are not being used as computers at all, when the theory has not been developed far enough for users to be confident that computational diagrams will commute. In these cases, users are experimenting on the devices, to develop their potential to act as computers, rather than using them to compute \emph{per se}. This is now the challenge to researchers in non-standard computing: to develop their device theory sufficiently that the elements within this framework are present in the system. Then it will be possible to argue with much more confidence than previously that the system is in fact performing computation.

The implications of this formalism go wider even than this. The framework shows the interaction of physical objects and the representations that we give them, in science, technology, and computing. By formalising this relationship, we now have a precise language in which to describe and understand how logical, mathematical, and computational structures interface with the physical objects of the world around us. This is the language of computing as standing on the boundaries between the physical theories of the underlying objects and interactions used, the technology that comes from engineering systems, and the mathematics and logic of the abstract computation. The study of physical computing has its own unique representation of physical systems and processes, and now has the foundational formalism in which to describe and determine its own domain, and its relation to the physics, chemistry, and biology of physical systems. Computer science takes its place as a natural science.\\

\noindent \emph{Acknowledgements} \ \ DH acknowledges very helpful discussion with James Ladyman, and also thanks members of the Oxford quantum nanoscience project for comments on an earlier version of this work. Our thanks to anonymous referees for valuable feedback and textual suggestions, and to Hacker News user csense for reference \cite{dogstick}. RCW was supported by the UK Engineering and Physical Sciences Council; VK was supported by a UK Royal Society University Research Fellowship. DH is supported by the CHIST-ERA DIQIP project, and by the FQXi Large Grant ``Time and the Structure of Quantum Theory". SS acknowledges partial funding  by the EU FP7 FET Coordination Activity TRUCE (Training and Research in Unconventional Computation in Europe), project reference number 318235.

\bibliographystyle{unsrt}
\bibliography{nsc}

\end{document}